\documentclass{aa}  
\usepackage{threeparttable}
\usepackage{natbib}
\usepackage{xcolor}
\usepackage{placeins}

\usepackage{microtype}
\usepackage{graphicx}
\usepackage{txfonts}
\PassOptionsToPackage{hyphens}{url}
\usepackage{hyperref}

\begin{document}

   \title{Diversity of nuclear star cluster formation mechanisms revealed by their star formation histories\thanks{Based on observation collected at the ESO Paranal La Silla Observatory,
Chile, Prog. ID 60.A-9192, 296.B-5054, 096.B-0399, 094.B-0895, 101.C-0329, 102.B-0455, 098.B-0619, and 100.B-0573.}}

   \author{K. Fahrion\inst{1}
          \and
          M. Lyubenova\inst{1}\
          \and
          G. van de Ven\inst{2} 
          \and
          M. Hilker\inst{1}
          \and
          R. Leaman\inst{2}
           \and
          J. Falc\'{o}n-Barroso\inst{3}\fnmsep\inst{4} 
          \and
          A. Bittner\inst{1}
          \and
           L. Coccato\inst{1}
           \and
           E. M. Corsini\inst{6}\fnmsep\inst{7}         
           \and
           D. A. Gadotti\inst{1}
          \and
          E. Iodice\inst{8}
           \and           
          R. M. McDermid\inst{9}\fnmsep\inst{10}\fnmsep\inst{11}
          \and
           I. Mart\'{i}n-Navarro\inst{3}\fnmsep\inst{4}
          \and 
          F. Pinna\inst{5}
          \and
          A. Poci\inst{10}
          \and
           M. Sarzi\inst{12}\fnmsep\inst{13}
          \and
          P. T. de Zeeuw\inst{15}\fnmsep\inst{16}
          \and
          L. Zhu\inst{17}}

   \institute{European Southern Observatory, Karl-Schwarzschild-Stra\ss{}e 2, 85748 Garching bei M\"unchen, Germany\\
   				\email{kfahrion@eso.org}
         \and 
             Department of Astrophysics, University of Vienna, T\"urkenschanzstrasse 17, 1180 Wien, Austria
             \and 
			Instituto de Astrof\'isica de Canarias, Calle Via L\'{a}ctea s/n, 38200 La Laguna, Tenerife, Spain
			\and 
			Depto. Astrof\'isica, Universidad de La Laguna, Calle Astrof\'isico Francisco S\'{a}nchez s/n, 38206 La Laguna, Tenerife, Spain
             \and 
             Max-Planck-Institut f\"ur Astronomie, K\"onigstuhl 17, 69117 Heidelberg, Germany
               \and 
                Dipartimento di Fisica e Astronomia 'G. Galilei', Universit\`a di Padova,  vicolo dell'Osservatorio 3, I-35122 Padova, Italy
			\and 
			INAF--Osservatorio Astronomico di Padova, vicolo dell'Osservatorio 5, I-35122 Padova, Italy
			 \and 
             INAF-Astronomical Observatory of Capodimonte, via Moiariello 16, I-80131, Napoli, Italy
               \and 
             Department of Physics and Astronomy, Macquarie University, North Ryde, NSW 2109, Australia
             \and 
             Astronomy, Astrophysics and Astrophotonics Research Centre, Macquarie University, Sydney, NSW 2109, Australia
             \and 
             ARC Centre of Excellence for All Sky Astrophysics in 3 Dimensions (ASTRO 3D), Australia
             \and 
             Armagh Observatory and Planetarium, College Hill, Armagh, BT61 9DG, Northern Ireland, UK
             \and 
             Centre for Astrophysics Research, University of Hertfordshire, College Lane, Hatfield AL10 9AB, UK
             \and 
             Sterrewacht Leiden, Leiden University, Postbus 9513, 2300 RA Leiden, The Netherlands
             \and 
             Max-Planck-Institut f\"ur Extraterrestrische Physik, Gie\ss{}enbachstraße 1, 85748 Garching bei M\"unchen, Germany
             \and 
             Shanghai Astronomical Observatory, Chinese Academy of Sciences, 80 Nandan Road, Shanghai 200030, China
             }

   \date{}

 
   \abstract{Nuclear star clusters (NSCs) are the densest stellar systems in the Universe and are found in the centres of all types of galaxies. They are thought to form via mergers of star clusters such as ancient globular clusters (GCs) that spiral to the centre as a result of dynamical friction
   or through \textit{in-situ} star formation directly at the galaxy centre.
   There is evidence that both paths occur, but the relative contribution of either channel and their correlation with galaxy properties are not yet constrained observationally.
   We aim to derive the dominant NSC formation channel for a sample of 25 nucleated galaxies, mostly in the Fornax galaxy cluster, with stellar masses between $M_\text{gal} \sim 10^8$ and $10^{10.5} M_\sun$ and NSC masses between $M_\text{NSC} \sim 10^{5}$ and $10^{8.5} M_\sun$.
   Using Multi-Unit Spectroscopic Explorer (MUSE) data from the Fornax 3D survey and the ESO archive, we derive star formation histories, mean ages and metallicities of NSCs, and compare them to the host galaxies.
  In many low-mass galaxies, the NSCs are significantly more metal-poor than their hosts with properties similar to GCs. In contrast, in the massive galaxies, we find diverse star formation histories and cases of ongoing or recent \textit{in-situ} star formation. Massive NSCs ($> 10^7 M_\sun$) occupy a different region in the mass-metallicity diagram than lower mass NSCs and GCs, indicating a different enrichment history.
   We find a clear transition of the dominant NSC formation channel with both galaxy and NSC mass. We hypothesise that while GC-accretion forms the NSCs of the dwarf galaxies, central star formation is responsible for the efficient mass build up in the most massive NSCs in our sample. At intermediate masses, both channels can contribute. The transition between these formation channels seems to occur at galaxy masses $M_\text{gal} \sim 10^9 M_\sun$ and NSC masses $M_\text{NSC} \sim 10^7 M_\sun$.}

   \keywords{galaxies: star clusters: general -- galaxies: clusters: individual: Fornax -- galaxies: nuclei
               }

   \maketitle
%
\section{Introduction}
\label{sect:intro}
Nuclear star clusters (NSCs) are dense star clusters located in the centres of a majority of galaxies. With stellar masses of 10$^5 - 10^8 M_\sun$ and effective radii of 3 - 20 pc \citep{Walcher2006, Georgiev2016, Spengler2017, Neumayer2020}, NSCs are even more massive and denser than globular clusters (GCs), making them the densest stellar systems in the Universe \citep{Walcher2005, Hopkins2010, Neumayer2020}. 
Due to their brightness, NSCs can be observed in galaxies beyond the Local Group and deep photometric studies in the Virgo and Fornax galaxy clusters ($D \approx$ 16 and 20 Mpc, respectively, see \citealt{Mei2007, Blakeslee2009}) have shown that the fraction of galaxies that host a NSC at their centre is a function of galaxy mass and reaches $> 90\%$ at galaxy masses of $\sim 10^9 M_\sun$ \citep{SanchezJanssen2019}. 

Located at the bottom of the potential well in a galaxy, NSCs are subject to various physical processes that shape the centres of galaxies. NSCs are known to co-exist with central supermassive black holes (SMBHs), as for example in the Milky Way (MW; \citealt{NeumayerWalcher2012, Georgiev2016, Nguyen2019, Neumayer2020}). However, interactions with SMBHs were suggested to destroy NSCs or inhibit NSC growth in the most massive galaxies \citep{Cote2006, NeumayerWalcher2012, Antonini2015, ArcaSedda2016}, possibly explaining the negligible nucleation fraction for galaxies with stellar masses $M_\text{gal} > 10^{11} M_\sun$. 

After the first studies found that NSC properties correlate with the properties of their host galaxy in a similar fashion to SMBHs \citep{Balcells2003, GrahamGuzman2003}, subsequent effort was put into establishing relations between the NSC mass and galaxy properties such as bulge luminosity, velocity dispersion, and total stellar mass \citep{Ferrarese2006, Wehner2006, Rossa2006}. 
It was suggested that NSCs and SMBHs follow the same scaling relations \citep{Ferrarese2006}, but larger sample sizes found different relations \citep{Erwin2012, Georgiev2016, Ordenes2018, SanchezJanssen2019}. For example, the slope of the NSC-galaxy mass relation is still debated. Studies of mostly massive galaxies found slopes $\sim 1$ \citep{Georgiev2016}, similar to the black hole-bulge mass relation \citep{McConnell2013, Saglia2016}, whereas recent works focusing on nucleated dwarf galaxies found shallower slopes indicating that this relation might be non-linear with a steeper slope for galaxy masses $\gtrsim 5 \times 10^9 M_\sun$ \citep{Ordenes2018, SanchezJanssen2019}. Also the $M - \sigma$ relation that connects the stellar velocity dispersion of a galaxy bulge to the mass of the central massive object, appears to differ between NSCs and SMBHs. While the $M_\text{BH} - \sigma$ relation has a slope between $\sim$ 4 and 5 \citep{KormendyHo2013, Saglia2016, CapuzzoDolcetta2017}, the $M_\text{NSC} - \sigma$ relation appears to be shallower, with a slope of $\sim$ 2 \citep{ScottGraham2013, CapuzzoDolcetta2017, Nguyen2018}.

The scaling relations between NSCs and their host galaxies suggest a connected evolution, but the formation mechanisms of NSCs are not yet completely understood, and there are two main pathways that are discussed (see review by \citealt{Neumayer2020}). NSCs (and nuclear stellar disks) might form from the subsequent merger of gas-free GCs that spiral inward due to dynamical friction 
\citep{Tremaine1975, CapuzzoDolcetta1993, CapuzzoDolcetta2008, Agarwal2011, Portaluri2013, ArcaSedda2014}. This scenario directly connects NSC formation to the GC population and is usually invoked to explain the 
presence of metal-poor stellar populations found especially in the NSCs of dwarf galaxies \citep{Rich2017, AlfaroCuello2019, Fahrion2020a, Johnston2020}.
Although the infall of GCs from different orbits can lead to low angular momentum in the NSC, detailed \textit{N}-body simulations of this scenario have shown that the resulting NSC can still have significant rotation \citep{Antonini2012, Perets2014, Tsatsi2017}. \cite{Lyubenova2019} explored the spatially resolved stellar kinematics of six galaxy nuclei hosting NSCs and compared these with the aforementioned $N$-body simulations of NSC formation. They concluded that, based on kinematics alone, the pure GC merging scenario cannot be excluded as a viable path for the formation of NSCs in intermediate to high-mass galaxy hosts. 

Alternatively, NSCs might form independently from the GC population via \textit{in-situ} star formation directly at the galaxy centre \citep{Loose1982, Milosavljevic2004, Schinnerer2007, Bekki2006, Bekki2007, Antonini2015}, explaining the young stars found in some NSCs of late-type galaxies, dwarf ellipticals and low-mass S0 galaxies \citep{Seth2010, Paudel2011, FeldmeierKrause2015, Carson2015, Kacharov2018, Nguyen2017, Nguyen2018, Paudel2020}. 
The \textit{in-situ} formation channel depends on internal feedback mechanisms and the available gas content. Many different mechanisms for funnelling gas to the centre have been studied. \cite{MihosHernquist1994} suggested that gas-rich mergers of galaxies will lead to a build up of gas in the merger remnant and the subsequent star formation can form compact objects similar to NSCs. Simulations by \cite{Hopkins2010a} showed that gas infall to the centre can happen as a result of gravitational torques during gas-rich mergers that allow gas clouds to collapse gravitationally, but there are also \textit{in-situ} NSC formation mechanisms that do not require gas-rich mergers. \cite{Milosavljevic2004} suggested that magneto-rotational instabilities in gas disks can funnel gas towards the nucleus with rates sufficiently for continuous star formation.

Although the GC-infall formation channel is generally focused on old GCs, it has to be noted that infalling clusters can also be younger massive clusters \citep{Agarwal2011}. \cite{ArcaSedda2016b} presented a $N$-body model of the dwarf starburst galaxy Henize 2-10 and found that the central star clusters of this galaxy will build a NSC within a few tens of Myr. A similar $N$-body simulation approach was used by \cite{Schiavi2021} to study the evolution of two star clusters found near the centre of the star-forming galaxy NGC\,4654. The authors concluded that these star clusters will merge to a NSC within 100 Myr. Using numerical simulations, \cite{Guillard2016} proposed a composite 'wet migration' scenario where a massive cluster forms in the early stages of galaxy evolution in a gas-rich disk and then migrates to the centre while keeping its initial gas reservoir, possibly followed by mergers with other gas-rich clusters. \cite{Paudel2020} identified off-centre young star clusters in dwarf elliptical galaxies and suggested that those could be examples of seed NSCs in such a wet-merger scenario. As these young massive clusters with ages $<$ 10 Myr can form close to the centres of their hosts (e.g. \citealt{Kornei2009, Nguyen2014, Georgiev2014}), they can also lead to young, metal-rich populations in the NSCs \citep{ArcaSedda2014, ArcaSedda2016, Schiavi2021}. Consequently, from stellar population analysis alone it is difficult to discern whether young and metal-rich populations are a result of star formation directly happening in the NSC or from clustered star formation in the central regions.

The NSC in the MW might be an example of both NSC formation processes acting together. Its star formation history (SFH) and the presence of young stars indicate formation from \textit{in-situ} star formation, but a recently discovered population of metal-poor stars might be explained by an infalling metal-poor GC \citep{FeldmeierKrause2020, ArcaSedda2020}. 
In general, the broad range of structural, kinematic, and stellar population properties of NSCs suggests that most likely both formation processes are realised in nature \citep{Hartmann2011, Antonini2013, Antonini2015, Guillard2016}. Although the observed $M_\text{NSC} - \sigma$ relation is in agreement with expectations from the GC accretion scenario \citep{Antonini2013, ArcaSedda2014, Gnedin2014, Antonini2015}, \cite{Neumayer2020} noted that the NSC-to-host mass scaling relation matches expectations from GC inspiral only for low-mass galaxies \citep{Gnedin2014, Ordenes2018, SanchezJanssen2019}.
Based on stellar population analysis of nucleated dwarf and massive early-type galaxies \citep{Koleva2009, Paudel2011, Spengler2017}, \cite{Neumayer2020} suggested that the dominant NSC formation channel switches from GC-accretion at low galaxy masses to \textit{in-situ} formation for massive galaxies. They proposed that this transition happens at galaxy masses of $M_\ast \sim 10^9 M_\sun$, where also the nucleation fraction of galaxies peaks and the NSC-galaxy mass relation steepens \citep{Ordenes2018, SanchezJanssen2019}. This is also supported by dynamical arguments, as the dynamical friction timescales for GCs to effectively spiral into the centres exceed the Hubble time in massive galaxies \citep{Turner2012, Antonini2013}.


Nonetheless, neither the relative contribution of either channel nor the dependence on host galaxy properties are fully understood. 
To date, solid observational evidence for a transition of the dominant NSC formation channel with galaxy properties is still lacking because of heterogeneous datasets, large uncertainties, and restricted sample sizes. In this paper, we present a spectroscopic analysis of a sample of 25 nucleated galaxies, 23 in the Fornax galaxy cluster and two in the Virgo cluster, that span galaxy masses between $10^8$ and $10^{10.5} M_\sun$ and NSC masses between $10^5$ and $10^{8.5} M_\sun$. 
We chose these galaxies due to the availability of Multi Unit Spectroscopic Explorer (MUSE, \citealt{Bacon2010}) data which was previously used in stellar population and GC system studies \citep{F3D_Survey, Pinna2019a, Pinna2019b, Iodice2019, Johnston2020, Fahrion2020a, Fahrion2020b}. We derive SFHs, ages, and metallicities of the host galaxies and their NSCs in a homogeneous fashion that enables a direct comparison of these components and allows us to observationally derive the dominant NSC formation channel as a function of galaxy properties. 
The data are described in Sect. \ref{sect:data}. Section \ref{sect:analysis} explains how the spectra of the different components are extracted and analysed, and the results are presented in Sect. \ref{sect:results}. We discuss the results in Sect. \ref{sect:discussion} and conclude in Sect. \ref{sect:conclusion}.

\section{Data}
\label{sect:data}
We used data acquired with the MUSE instrument on ESO's Very Large Telescope (VLT) in Chile. MUSE is an integral-field spectrograph that provides a $1\arcmin\times1\arcmin$ field of view, sampled at 0.2\arcsec pixel$^{-1}$. In the wavelength dimension, MUSE covers the optical regime from 4700 to 9300 \AA, sampled at 1.25\,\AA\ pixel$^{-1}$, with a mean full width at half maximum (FWHM) of the line spread function of $\sim$ 2.5 \AA. 

We used MUSE data from the Fornax3D survey (F3D; \citealt{F3D_Survey}), a magnitude limited survey that targets bright galaxies in the Fornax galaxy cluster ($D \approx 20$ Mpc, \citealt{Blakeslee2009}). For details concerning the data reduction, we refer to \cite{F3D_Survey} and \cite{Iodice2019}. Spectroscopic GC catalogues derived from this data were presented in \cite{Fahrion2020b, Fahrion2020c}. In the present work, we concentrated on the galaxies that are defined as nucleated based on the analysis of \textit{Hubble Space Telescope} data of the Advanced Camera for Survey (ACS) Fornax Cluster Survey (ACSFCS, \citealt{Jordan2007}) by \cite{Turner2012}. 

In addition to the F3D galaxies, we complemented the sample with Fornax dwarf galaxies that were also in the sample analysed in \cite{Johnston2020}. We included these galaxies to extend the mass range of studied galaxies to $< 10^9 M_\sun$. Because our study relies on an homogeneous analysis of all galaxies using the same techniques to allow us a reliable comparison between galaxies, we included these galaxies in our analysis. Furthermore, we included two nucleated dwarf galaxies in the Virgo cluster (VCC\,990 and VCC\,2019) that were previously analysed by \cite{Bidaran2020} with focus on their internal kinematics. For the Virgo and Fornax dwarf galaxies, we downloaded the reduced MUSE data cubes from the ESO science archive and applied the Zurich Atmosphere Purge (ZAP) algorithm \citep{Soto2016} to further reduce sky residuals. For FCC\,202 we used the two pointings independently, because the combined mosaic provided by the archive showed a sub-optimal co-addition and the NSC is only covered in one.

Additionally, we added FCC\,47, an early-type galaxy outside of the virial radius of the Fornax cluster, for which we acquired adaptive-optics supported MUSE science verification data in 2017. FCC\,47 and its massive NSC were analysed in detail in \cite{Fahrion2019a, Fahrion2019b}, but we include it here to ensure a homogeneous comparison to the rest of the sample. 

Table \ref{tab:galaxy_overview} gives an overview of the observed galaxies. The galaxies comprise early-type galaxies and dwarf ellipticals. 
In this table, we give the references for the galaxy and NSC masses. Where no NSC masses were available in the literature, we derived them from their ($g - z$) colours from the ACSFCS \citep{Turner2012} and ACS Virgo Cluster Survey \citep{Cote2006} and the photometric predictions of the E-MILES stellar population synthesis models \citep{Vazdekis2016} that give the mass-to-light ratio in the ACS $g$-band. To determine the mass, we sample the $g$ and $z$ magnitudes within their uncertainties and allow ages between 5 and 14 Gyr. The age can affect the mass-to-light ratio, but the largest uncertainty stems from the uncertainties of the NSC $g$ and $z$ magnitudes ($\sim 0.2$ mag).

\begin{table*}[]
    \centering
    \caption{Overview of nucleated galaxies.}
    \begin{threeparttable}
    \begin{tabular}{l l c c c c c c}\hline
       Galaxy  & Altern. name  & $R_\text{eff}$ & $R_\text{eff}$ & log($M_\text{gal}$) & log($M_\text{NSC}$) & Exp. time & Programme \\
            &   & (arcsec) & (kpc) &  &  & (sec) \\
       (1) & (2) & (3) & (4) & (5) & (6) & (7) & (8)\\
       \hline\hline
        FCC\,47  & NGC\,1336 & 30.0\tnote{a} & 2.9 & 9.97\tnote{d} & 8.74\tnote{h} & 3600 & 060.A-9192 (PI: Fahrion) \\
         FCC\,119  &  -- & 17.4\tnote{b} & 1.7 & 9.0\tnote{d} & 6.81\tnote{h} & 5400 & 296.B-5054 (PI: Sarzi) \\
        FCC\,148 & NGC\,1375 & 28.3\tnote{b} & 2.7 & 9.76\tnote{d} &  8.37\tnote{h} & 3600 + 5400 & 296.B-5054 (PI: Sarzi)\\
        FCC\,153 & IC\,1963 & 19.8\tnote{b} & 1.9 & 9.88\tnote{d} &  7.29\tnote{h} & 3600 + 5400 & 296.B-5054 (PI: Sarzi)\\
        FCC\,170  & NGC\,1381 & 15.9\tnote{b} & 1.5 & 10.35\tnote{d}  &  8.45\tnote{h} & 3600 + 7200 & 296.B-5054 (PI: Sarzi)\\
        FCC\,177 & NGC\,1380A & 35.9\tnote{b} & 3.5 & 9.93\tnote{d}  &  7.83\tnote{h} & 3600 + 5400 & 296.B-5054 (PI: Sarzi)\\
        FCC\,182  & -- &  9.9\tnote{b} & 1.0 &  9.18\tnote{d}  &  6.03\tnote{e} & 5400 &  296.B-5054 (PI: Sarzi)\\
        FCC\,188  & -- &  12.1\tnote{a} & 1.2 &  8.89\tnote{e}  &  6.85\tnote{e} & 10400 & 096.B-0399 (PI: Napolitano)\\
        FCC\,190 & NGC\,1380B & 18.3\tnote{b} & 1.8 & 9.73\tnote{d}  &  7.18\tnote{h} & 3600 + 5400 & 296.B-5054 (PI: Sarzi)\\
        FCC\,193 & NGC\,1389 & 28.2\tnote{b} & 2.7 & 10.52\tnote{d}  &  8.15\tnote{h} & 3600 + 5400 & 296.B-5054 (PI: Sarzi)\\
        FCC\,202 & NGC\,1396 & 9.8\tnote{a} & 1.0 & 9.03\tnote{e}  &  6.76\tnote{e} & 10400 + 10400 & 094.B-0895 (PI: Lisker)\\
        FCC\,211 & -- &  5.6\tnote{a} & 0.5 & 8.52\tnote{e} &  6.70\tnote{e} & 10400 & 096.B-0399 (PI: Napolitano)\\
        FCC\,215 & -- &  7.4\tnote{a} & 0.7 & 6.79\tnote{e}  &  5.94\tnote{e} & 10400 & 096.B-0399 (PI: Napolitano)\\
        FCC\,222 & AM\,0337-353 & 14.5\tnote{a} & 1.4 &  8.80\tnote{e}  &  6.45\tnote{e} & 10400 & 096.B-0399 (PI: Napolitano)\\
        FCC\,223 & AM\,0337-355 & 16.6\tnote{a} & 1.6 & 8.78\tnote{e}  &  6.38\tnote{e} & 10400 & 096.B-0399 (PI: Napolitano)\\
        FCC\,227 & -- & 7.3\tnote{a} & 0.7 & 6.73\tnote{e} & 5.24\tnote{e} & 10400 & 096.B-0399 (PI: Napolitano)\\
        FCC\,245  & -- & 12.9\tnote{a} & 1.3 & 8.77\tnote{e}  &  6.05\tnote{e} & 1350 & 101.C-0329 (PI: Vogt)\\
        FCC\,249 &  NGC\,1419 & 9.6\tnote{b} & 0.9 &  9.70\tnote{d}  &  6.93\tnote{h} & 3600 + 1800 & 296.B-5054 (PI: Sarzi)\\
        FCC\,255 & ESO\,358-G50 & 13.8\tnote{b} &  1.3 & 9.70\tnote{d}  &  6.98\tnote{h} & 3600 + 1800 & 296.B-5054 (PI: Sarzi)\\
        FCC\,277 & NGC\,1428 &  12.8\tnote{b} & 1.2 & 9.53\tnote{d}  &  7.22\tnote{h} & 3600 + 2160 & 296.B-5054 (PI: Sarzi)\\
        FCC\,301 & ESO 358-G59 & 11.7\tnote{b} & 1.1 & 9.30\tnote{d}  &  6.91\tnote{h} & 3600 + 1800 & 296.B-5054 (PI: Sarzi)\\
        FCC\,310 & NGC\,1460 &  35.6\tnote{b} & 3.5 & 9.73\tnote{d}  &  7.81\tnote{h} & 3600 + 5400 & 296.B-5054 (PI: Sarzi)\\
        FCCB\,1241 & AM\,0336-323 &  8.4\tnote{b} & 0.8 & 8.13\tnote{e}  &  5.48\tnote{e} & 2160 & 102.B-0455 (PI: Johnston) \\
        VCC\,990  & IC\,3369 & 8.1\tnote{c} & 0.6 & 9.10\tnote{f} & 6.83\tnote{h} & 3240 & 098.B-0619 (PI: Lisker) \\
        VCC\,2019 & IC\,3735 & 15.7\tnote{c} & 1.3 & 9.01\tnote{g} & 6.78\tnote{g} & 3744 & 100.B-0573 (PI: Lisker) \\ \hline
    \end{tabular}
    \begin{tablenotes}
    \item (1, 2) galaxy name and alternative name, (3, 4) effective radius in arcsec and in kpc assuming a distance of 20 Mpc (Fornax, \citealt{Blakeslee2009}) and 16.5 Mpc (Virgo, \citealt{Mei2007}) (5) stellar mass of the host galaxy, (6) stellar mass of the NSC (7) total exposure time of the pointings covering the galaxy. (8) MUSE observing programme and PI. References: a - \cite{Ferguson1989}, b - \cite{Iodice2019}, c - \cite{Kim2014}, d - \cite{Liu2019}, e - \cite{Johnston2020}, f - \cite{Peng2008}, g - \cite{Paudel2011}, h - this work.
    \end{tablenotes}
    \end{threeparttable}
    \label{tab:galaxy_overview}
\end{table*}

\section{Analysis}
\label{sect:analysis}
Understanding the formation of NSCs requires a comparison of the properties of the NSC to the host galaxy from its central regions to the outskirts. In the following, we describe how we extracted the MUSE spectra of the different spatial regions we aim to compare (see Table \ref{tab:spectra_overview} for an overview). Sect. \ref{sect:ppxf} then details how the spectra are fitted to extract the stellar population properties.

\subsection{Nuclear star cluster spectra}
\label{sect:NSC_analysis}
The biggest challenge to extract the NSC properties from MUSE data is to disentangle the NSC spectra from the underlying galaxy light. The spectrum obtained from a central aperture contains significant contributions from both the NSC and the galaxy. In addition, at the distances of the Fornax and Virgo galaxy clusters (20 Mpc and 16 Mpc, respectively), NSCs are not resolved. To reduce the contribution of the galaxy, we took the following approach that aims to subtract a spectrum representing the galaxy from the central spectrum. 

The NSC spectrum is given by
\begin{equation}
S_\text{NSC} = S_\text{central} - S_\text{host, central},
\end{equation}
where $S_\text{central}$ is the central spectrum acquired directly from the data. To extract this spectrum, we used a circular aperture weighted by the point spread function (PSF, usually at FWHM of 0.8\arcsec = 4 pixel) centred on the galaxy centre, assuming an unresolved NSC. $S_\text{host, central}$ is the (unknown) contribution of the host galaxy at this central position.

We estimate this galaxy contribution as
\begin{equation}
S_\text{host, central} = \alpha \, S_\text{host, 2\arcsec},
\end{equation}
where $S_\text{host, 2\arcsec}$ is the spectrum of the host galaxy at 2\arcsec defined by an elliptical annulus aperture with 8 pixel (1.6\arcsec = 200 pc in Fornax and 160 pc in Virgo) inner and 13 pixel (2.6\arcsec) outer radius. We chose this radius to determine the spectrum of the galaxy right outside the extend of the central PSF and thus outside the influence of the NSC. The position angle and ellipticity of the elliptical aperture were derived from a Multi Gaussian Expansion fit \citep{Emsellem1994, Cappellari2002} to the galaxy image. For the F3D galaxies, these are comparable to the ones listed in \cite{Iodice2019}. We assume that this spectrum represents the galaxy at the centre (and hence the stellar populations), but we allow for an additional scaling factor $\alpha$ that describes the flux level difference between the central and the annulus apertures:
\begin{equation}
    \alpha = \frac{F_\text{host, central}}{F_\text{host, 2\arcsec}}.
\end{equation}
As all fluxes are normalised by the extraction area, $\alpha = 1$ if the flux of the host galaxy is constant between 2\arcsec\,and the centre.

\begin{figure*}
    \centering
    \includegraphics[width=0.57\textwidth]{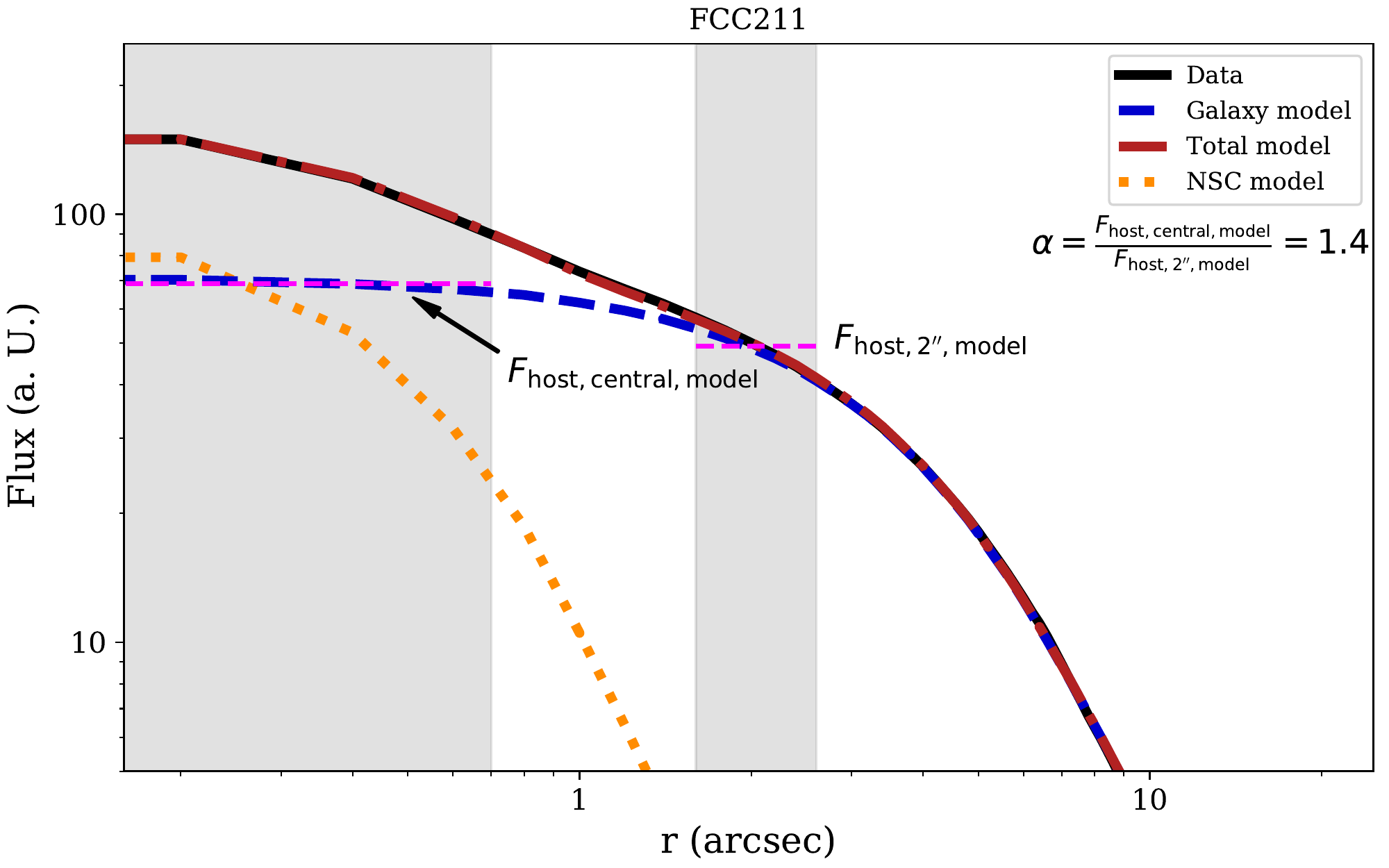}
    \includegraphics[width=0.37\textwidth]{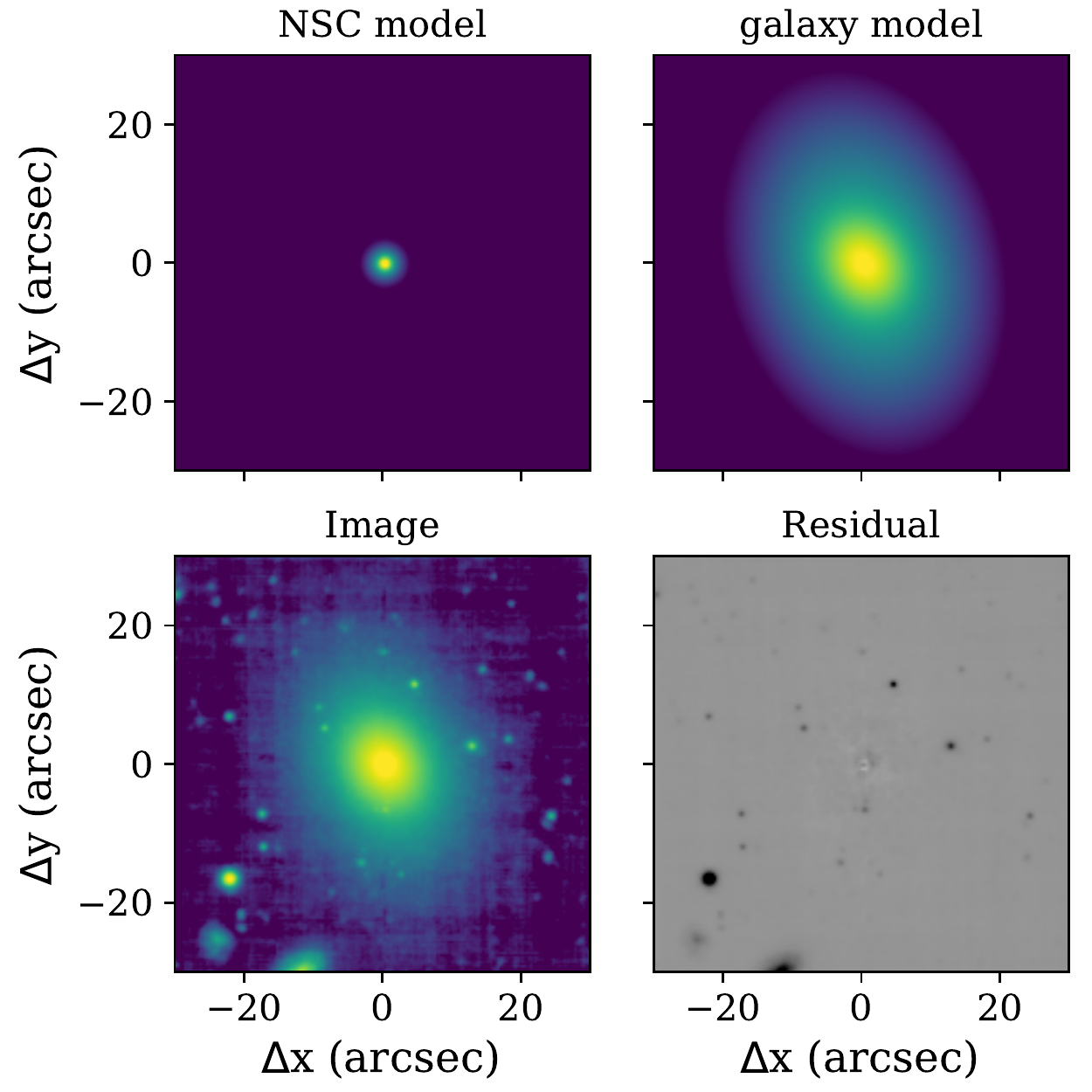}
    \caption{\textsc{imfit} modelling of FCC\,211 to decompose the white-light image into a galaxy and a NSC component. \textit{Left}: radial profile of the data (solid black line), galaxy model (dashed blue line), NSC model (dotted orange line), and combined model (dashed-dotted red line). The grey areas show the extraction regions. The mean fluxes in these regions are indicated by the horizontal dashed magenta lines to illustrate the scaling factor $\alpha$. \textit{Right}: 2D cutout images of the original image (bottom left), NSC and galaxy model (top row) and resulting residual image (bottom right). As described in the text, this decomposition is used to determine the contribution of galaxy light when extracting the NSC spectrum from the MUSE data. Each image shows a region of 60\arcsec $\times$ 60\arcsec (5.4 kpc $\times$ 5.4 kpc) centred on the NSC.}
    \label{fig:imfit_modelling}
\end{figure*}

In the dwarf galaxies, the NSC often dominates the central regions and for those, $\alpha = 1$ is a reasonable approximation. However, to optimise the NSC spectrum extraction, we used two-dimensional surface brightness modelling with \textsc{imfit} \citep{Erwin2015} to determine $\alpha$ on model images (see Fig. \ref{fig:imfit_modelling} for an illustration).
\textsc{imfit} is a modular 2D fitting procedure that fits images of galaxies with an user-specified set of 2D image functions (e.g. exponential, S\'ersic, Gaussian etc.). An arbitrary number of different components can be added together to model galaxy images. To fit the white light images of our MUSE data, we modelled the unresolved NSCs with a single component described by a Moffat function. A S\'ersic function yields comparable results. The galaxies, on the other hand, can consist of multiple components to capture the various physical components in the galaxy (disks, bulges etc.).

After decomposing the white-light MUSE images in this way, we obtained model images of the NSC and the host galaxies with the same spatial scales as the original MUSE data. Using these model images, we can apply the same apertures as used on the MUSE data to estimate $\alpha$:
\begin{equation}
    \alpha = \frac{F_\text{Host, central, model}}{F_\text{Host, 2\arcsec, model}}.
\end{equation}

This approach was inspired by the technique described in \cite{Johnston2020}, where the authors use similar 2D surface brightness modelling at every wavelength bin. In this way, they obtain single model spectra for every component. However, spatial stellar population gradients cannot be captured with this approach as every component is assigned a single spectrum. Dwarf galaxies, as those studied in \cite{Johnston2020}, usually lack strong stellar population gradients and thus this approach is completely valid. Our sample comprises more complex galaxies with multiple structural components and strong radial gradients (see Fig. \ref{fig:radial_profiles}). Therefore, we adopted the approach described here that allows us to base the spectra extraction on the MUSE data directly without invoking full galaxy models. We only used the \textsc{imfit} models to infer the scaling factor $\alpha$ and extracted the spectra of each spatial region directly from the data.

Figure \ref{fig:imfit_modelling} illustrates our approach applied to FCC\,211, one of the dwarf galaxies. The galaxy component is modelled in this case with two S\'ersic functions, while the NSC is described by a Moffat function. For this galaxy, we find $\alpha = 1.4$, implying that the contribution of the galaxy to the NSC spectrum is roughly 40\% higher than at the extraction radius of 2\arcsec. In general, we find values of $\alpha$ between 1.0 and 3.0. 
While the surface brightness distributions of dwarf galaxies in the sample can be modelled without strong residuals, some of the more massive galaxies in the sample have complex structures such as disks and X-shaped bulges which are not completely captured by the model. Nonetheless, the radial surface brightness profiles are well fitted also in these galaxies. 
In App. \ref{app:imfit}, we present the \textsc{imfit} model for FCC\,170, a massive edge-on S0 galaxy with a X-shaped bulge as well as a thin and thick stellar disk. The 2D residual of this galaxy (Fig. \ref{fig:imfit_modelling_FCC170}) still shows left-over structure, despite a good fit to the radial surface brightness profile. We find $\alpha = 2.8$ for this galaxy, but as discussed in App. \ref{app:imfit}, the uncertainty on $\alpha$ can be up to $\sim 20 \%$ in these massive galaxies as it depends on the subjective choice of the number of used components.


\subsection{Galaxy spectra}
\label{sect:gal_specs}
To compare the galaxies to their NSCs, we extracted averaged spectra of the galaxies at different radii. In addition to the region at 2\arcsec\,that was used to estimate the galaxy stellar populations at the centre, we further extracted averaged spectra at 0.5 and 1.0 effective radii ($R_\text{eff}$) using elliptical annulus apertures with 20 and 40 pixel width, respectively, and the same ellipticity as before. For some of the dwarf galaxies with small $R_\text{eff}$, we decreased the width of the extraction aperture to ensure that the NSC is excluded.
In addition, bright foreground stars and background galaxies are masked.
For FCC\,119, FCC\,188, FCC\,223, and FCC\,227, only the spectra obtained around 0.5 $R_\text{eff}$ yield a sufficient signal-to-noise ratio ($S/N$) for fitting due to the low surface brightness of these galaxies. 

\begin{figure*}
    \centering
    \includegraphics[width=0.95\textwidth]{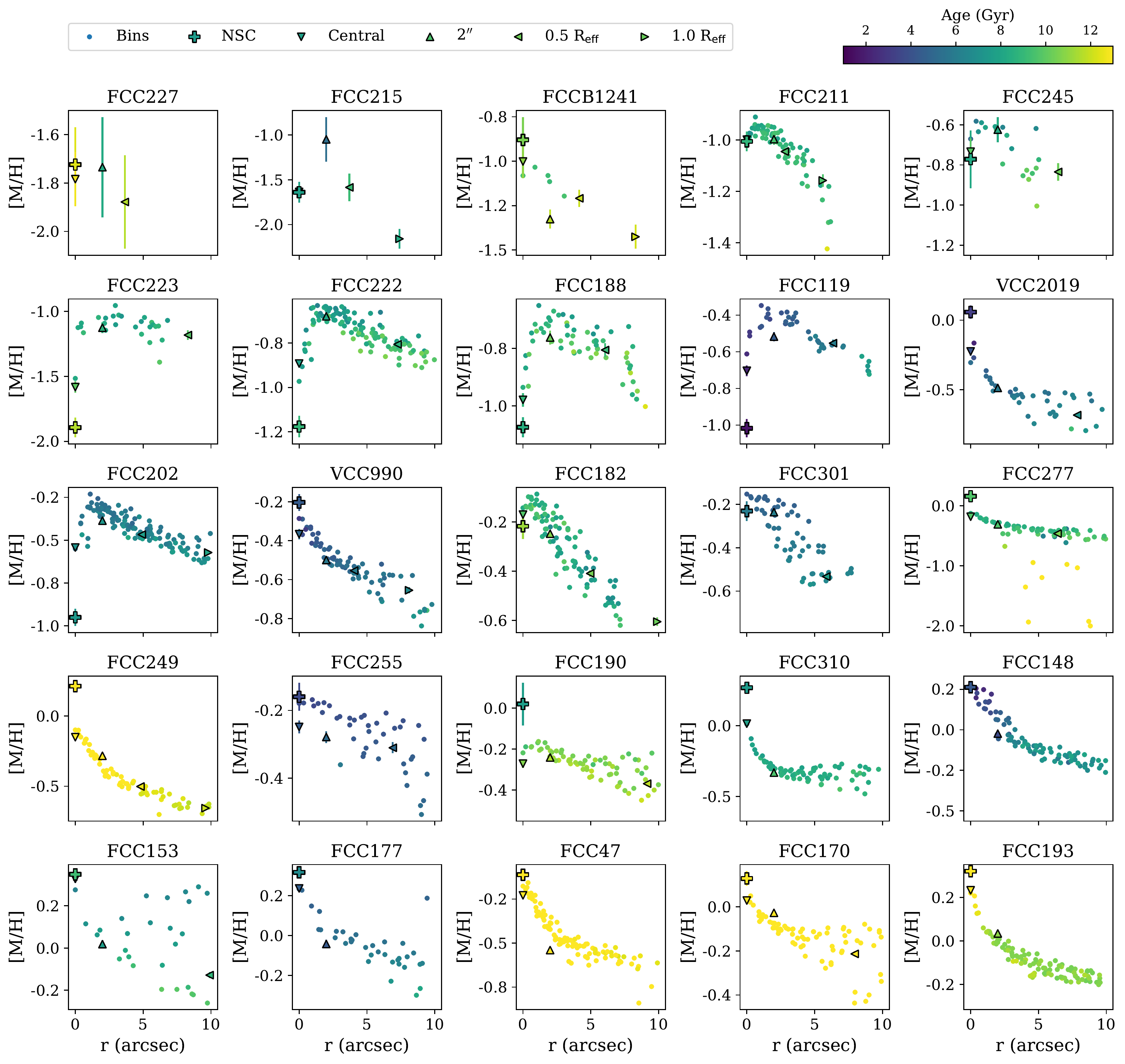}
    \caption{Radial metallicity profiles for the galaxies. Small dots show individual bins from the Voronoi binned cubes, crosses refer to the NSCs, triangles to the different extraction radii as described in the legend and Table \ref{tab:spectra_overview}. All symbols are coloured by age. Galaxies are ordered by increasing stellar mass from top left to bottom right. FCC\,227 and FCC\,215 are too faint to be binned.}
    \label{fig:radial_profiles}
\end{figure*}

To illustrate that these averaged spectra represent the stellar populations of the galaxies, we compare them to binned spectra obtained using the Voronoi binning routine described in \cite{Cappellari2003} in Fig. \ref{fig:radial_profiles}. This routine bins the MUSE data to a target $S/N$ and thus enables a continuous view of the stellar population properties. As we are interested in a comparison of the central galaxy regions to their NSCs, we chose the target $S/N$ such that there are sufficient bins in the inner 10\arcsec of each galaxy. In the most massive galaxies, this was reached for even very high target $S/N$ > 300, but for some of the dwarf galaxies, target values of $S/N = 30$ were chosen to ensure robust fits. The data of FCC\,227 and FCC\,215 are too shallow for binning. We show four metallicity maps as examples in App. \ref{app:maps} (Fig. \ref{fig:maps_zoom}).

\begin{figure*}[!htbp]
    \centering
    \includegraphics[width=0.97\textwidth]{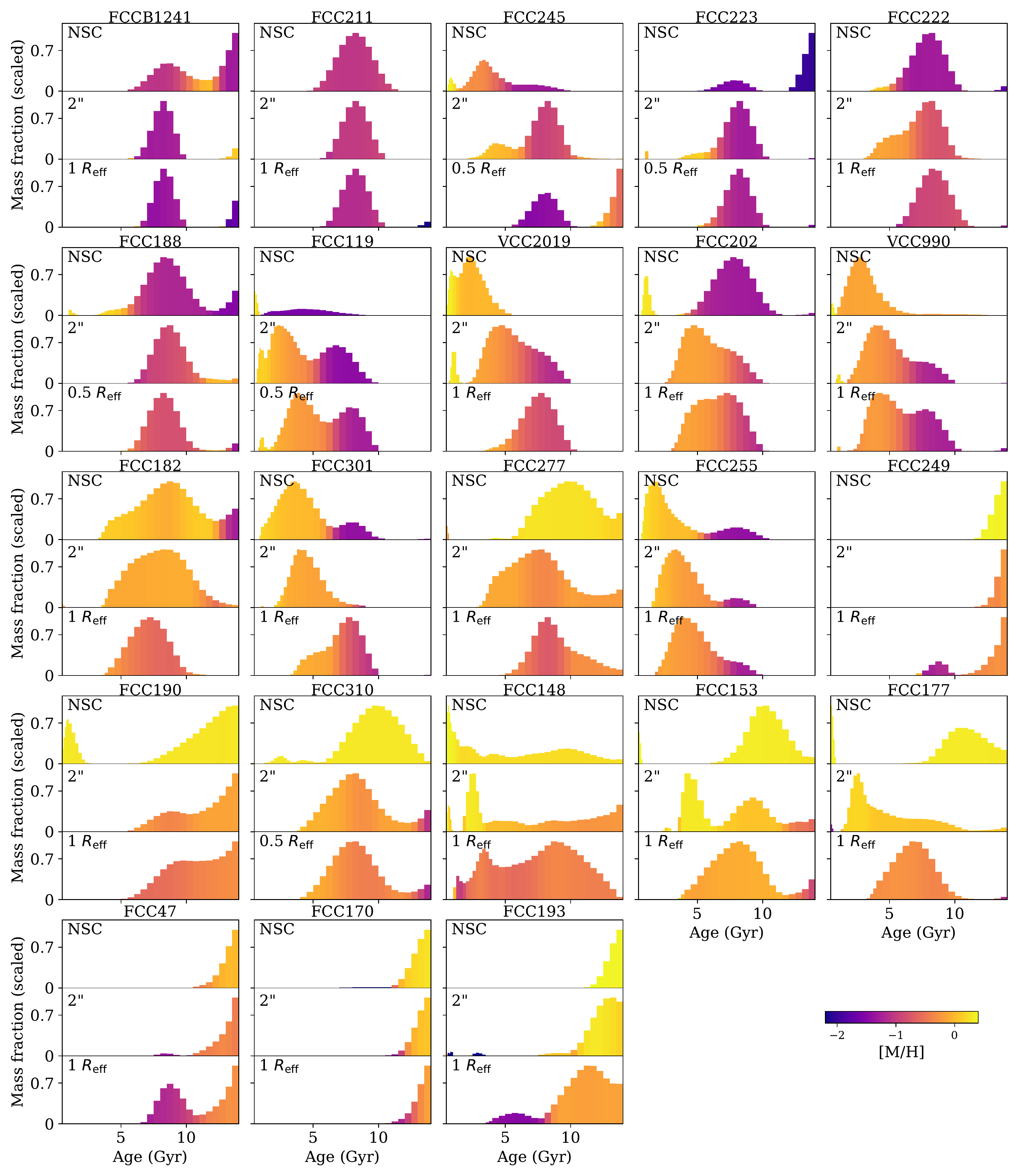}
    \caption{Star formation histories of NSCs (top), at the 2\arcsec\,extraction annulus (middle), and host galaxy (bottom), each panel corresponding to one galaxy. The mass fractions are scaled such that the peak is at 1.0 for visualisation of their shape. The galaxies are ordered by increasing stellar mass and the bins are colour-coded by their mean metallicity.}
    \label{fig:SFHs_coloured}
\end{figure*}

Table \ref{tab:spectra_overview} gives a summary of the different spectra and Fig. \ref{fig:radial_profiles} illustrates that the stellar population properties determined from these different annuli spectra are consistent with the properties obtained from the binned spectra. In general, we find that the stellar population results presented in the next sections are not sensitive to the choice of $\alpha$. As Fig. \ref{fig:radial_profiles} illustrates, the overall trends with metallicity are found even without subtracting the galaxy contribution. With this subtraction the respective trends are enhanced, illustrating that the NSC is responsible for the behaviour in the central bins. In contrast, the subtraction of the galaxy contribution has a stronger effect on the SFHs presented in Fig. \ref{fig:SFHs_coloured}, but here testing of our method has shown that similar SFHs are recovered when simply assuming $\alpha = 1$, possibly because our best-fit values for $\alpha$ reach moderate values between $1$ and $3$. Using unrealistically large values of $\alpha > 5$, however, can increase the associated uncertainties drastically and might even lead to a loss of the NSC signal.

\begin{table}[]
    \centering
        \caption{Overview of the different spectra.}
    \begin{tabular}{l l}
        \hline
        Name & Description \\ \hline
        Central & spectrum of the central PSF \\
        NSC & NSC spectrum (galaxy contribution subtracted) \\
        2\arcsec & annulus spectrum at 2\arcsec \\
        0.5 $R_\text{eff}$ & annulus spectrum at 0.5 $R_\text{eff}$ \\
        1.0 $R_\text{eff}$ & annulus spectrum at 1.0 $R_\text{eff}$ \\ 
        Bins & Voronoi-binned spectra \\\hline
    \end{tabular}

    \label{tab:spectra_overview}
\end{table}




\subsection{Full spectral fitting}
\label{sect:ppxf}
We fitted all spectra using full spectrum fitting with the penalised pixel-fitting routine \textsc{pPXF}, \citep{Cappellari2004, Cappellari2017} - a well established code that fits input spectra with a linear combination of user-provided template spectra. We used the single stellar population (SSP) model spectra from the extended Medium resolution INT Library of Empirical Spectra (E-MILES, \citealt{Vazdekis2010, Vazdekis2016}).
The E-MILES models with BaSTi isochrones \citep{Pietrinferni2004, Pietrinferni2006} give a grid of ages and total metallicities between 30 Myr and 14 Gyr and [M/H] = $-2.27$ dex and $+0.40$ dex, respectively. These so-called baseFe models inherit the abundance pattern of the MW due to their construction from empirical spectra. They have [Fe/H] = [M/H] at higher metallicities, but include $\alpha$-enhanced spectra at lowest metallicities.

We used a MW-like double power-law (bimodal) initial mass function (IMF) with a high mass slope of 1.30 \citep{Vazdekis1996}. The model spectra have a spectral resolution of 2.51\,\AA\, in the wavelength region between 4700 and 7100 \AA\, that we used in the fits \citep{FalconBarroso2011}, approximately corresponding to the mean instrumental resolution of MUSE ($\sim 2.5\,\AA)$. To account for the varying instrumental resolution of MUSE, we used the description of the MUSE line spread function from \cite{Guerou2016}. In FCC\,119 and FCC\,148, where emission lines from ionised gas are clearly visible in the spectra, we fitted them simultaneously. Telluric lines and remnant sky residual lines are masked in the fits. As recommended for example by \cite{Cappellari2017}, we first fitted for the line-of-sight (LOS) velocity distribution using additive polynomials of degree 12. Those ensure a smooth continuum and are required to find the best-fitting LOS velocity and velocity dispersion. To then fit for the stellar population parameters (age and metallicity), we kept the LOS velocity distribution fixed and used multiplicative polynomials of degree 8 instead of additive polynomials to preserve the line profiles.

\textsc{pPXF} returns the weights of the models used in the fit, which allows us to determine the mean age and metallicity as the weighted means. Because the E-MILES models are normalised to 1 $M_\sun$, the resulting stellar population properties are mass-weighted. 
To extract SFHs and metallicity distributions, we applied regularisation during the fit with \textsc{pPXF}. Regularisation leads to a smooth distribution of mass weights of the assigned models and hence allows to reconstruct smooth distributions of ages (the SFHs) and metallicities. To derive the regularisation parameter, we followed the established approach recommended by \cite{Cappellari2017} and \cite{McDermid2015}: The regularisation parameter is given as the one that increases the $\chi^{2}$ of the unregularised fit by $\sqrt{2\,N_{\text{pix}}}$, where $N_\text{pix}$ is the number of fitted spectral pixels. We calibrated this regularisation for the NSC spectrum of each galaxy. 

Using the calibrated regularisation parameter in \text{pPXF} returns the smoothest SSP model weight distribution that is still compatible with the data, meaning that not every small feature in the SFHs is related with a real star formation episode. Moreover, determining the correct regularisation parameter is a non-trivial task \citep{Cappellari2017, Pinna2019a, Boecker2019}. However, the overall trends and presence of multiple populations are reliable. We verified that the SFH distributions are robust against changes to the library of SSP models (e.g. using the scaled solar MILES models), order of the multiplicative polynomial, chosen wavelength range (e.g. extending to 8900\AA), and regularisation parameter. In our tests, the mean values of age and metallicity are recovered within their uncertainties and the general shape of the SFHs and metallicity distributions are preserved under these changes. This also holds for different choices of $\alpha$.

In addition to using regularisation, we also fitted the spectra with a Monte-Carlo (MC) approach to derive realistic random uncertainties (see also  \citealt{Cappellari2004, Bittner2019, Fahrion2020b}). After the first unregularised fit of the original spectrum, we used the corresponding residual (best-fit spectrum subtracted from the input spectrum) to perturb the best-fit spectrum randomly in each wavelength bin. This approach is repeated 100 times to create 100 realisations of the best-fit spectrum that are fitted independently. The resulting properties such as the LOS velocity, age, or metallicity are then obtained as the mean of the resulting distribution and the uncertainties are given by the standard deviation assuming a Gaussian distribution. We verified that the mean of the age and metallicity MC fits are consistent with the weighted means from the regularised fits, as shown in Appendix \ref{app:approaches}.

\begin{figure*}[!htbp]
    \centering
    \includegraphics[width=0.97\textwidth]{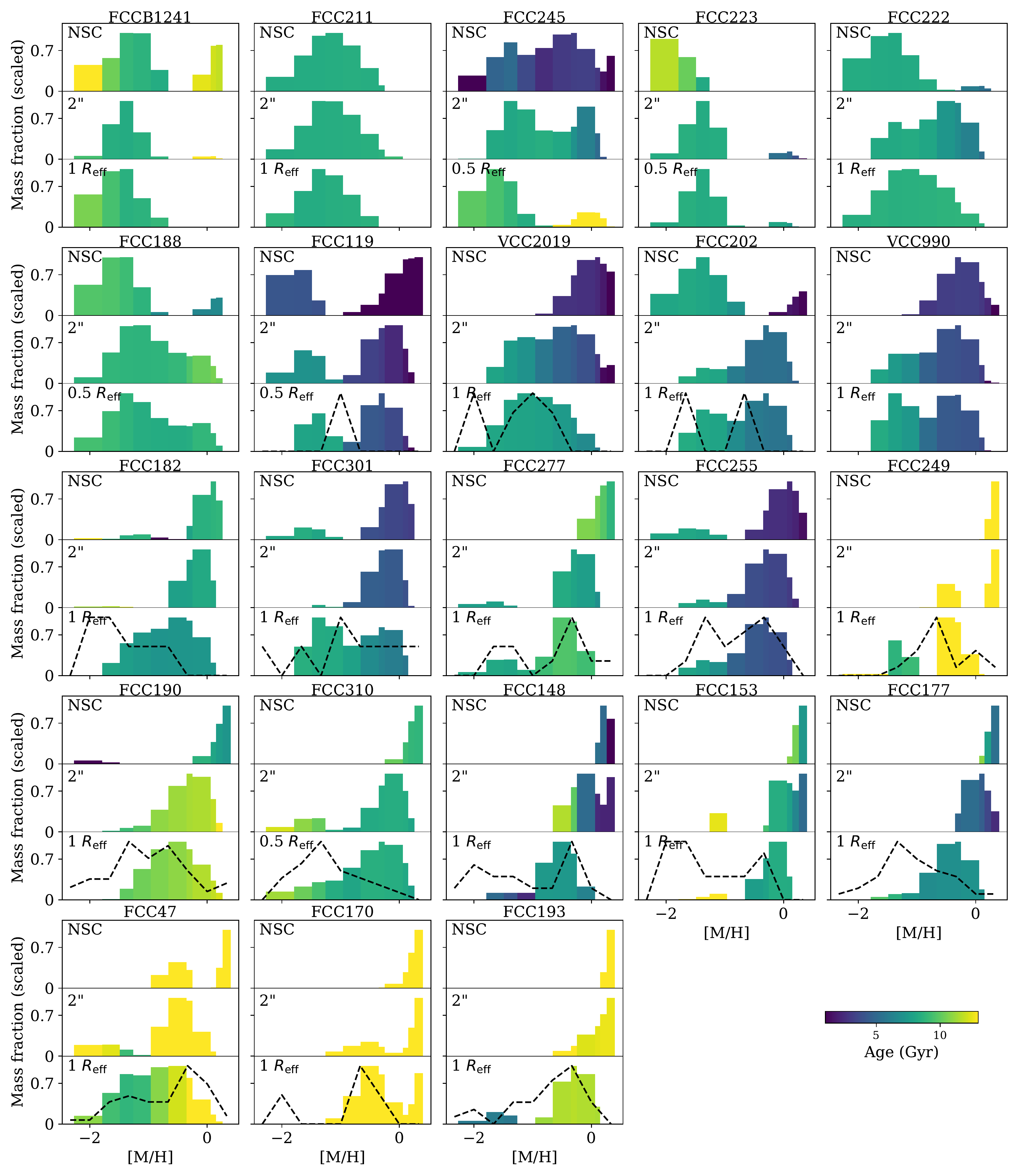}
    \caption{As in Fig. \ref{fig:SFHs_coloured} but for metallicity distributions. Dotted lines show the expected GC metallicity distributions. To obtain those, we applied the GC colour-metallicity relation from \cite{Fahrion2020c} to photometric GC catalogues from \cite{Jordan2007, Jordan2015}. For visualisation, the GC distributions are normalised to match the peak height for the galaxy mass fractions.}
    \label{fig:MDFs_coloured}
\end{figure*}

\section{Results}
\label{sect:results}
The following section presents the results from the stellar population analysis. First, we discuss the radial metallicity profiles and present the star formation histories and metallicity distributions. Sect. \ref{sect:MZR} then presents the mass-metallicity relations. Table \ref{tab:results} gives the resulting ages and metallicities for the NSCs and their host galaxies. 

\subsection{Radial metallicity profiles}
\label{sect:radial_profiles}
Figure \ref{fig:radial_profiles} shows the radial metallicity profiles obtained from the different spatial regions that we compare as well as from the Voronoi-binned data cubes. The profiles from the binned data were extracted using elliptical isophotes with same the position angle and ellipticity as used in Sect. \ref{sect:NSC_analysis}.
FCC\,227 and FCC\,215 are too faint to be binned. This figure displays that results obtained from the binned cubes are consistent with those obtained from the spectra extracted with different aperture sizes. The metallicities and ages determined from the spectra obtained at 2\arcsec, 0.5, and 1.0 $R_\text{eff}$ follow those from the binned data at the respective positions. In contrast, the NSC metallicities (obtained from the background-cleaned NSC spectra), follow the trends seen in the binned data, but can be more extreme - either more metal-poor or more metal-rich - than what is found in the central bins or the central spectrum. This is expected because the central binned spectra are affected by both the unresolved NSCs and the underlying galaxy, which dilutes the signal of the NSC. 

The binned radial metallicity profiles generally show an increase in metallicity from the outskirts to the central region ($< 2$\arcsec), as found for most galaxies (e.g. \citealt{Kuntschner2010, MartinNavarro2018, Zhuang2019, Santucci2020}). In the massive galaxies (bottom two rows), the highest metallicities are found in the central bins and their NSCs even surpass these metallicities. In contrast, many of the dwarf galaxies (top rows) show a drop in metallicity at the very centre (see also Fig. \ref{fig:maps_zoom}). As the comparison to the NSC metallicities shows, these central decreases are caused by metal-poor NSCs that dominate the light in the central bins of the dwarf galaxies. Except for FCC\,119, no central star formation is evident in these dwarf galaxies that could bias the results and for FCC\,119 we find similar results when masking the emission lines and fitting the full MUSE wavelength range up to 8900 \AA.

\begin{table*}[]
    \centering
    \caption{Mass-weighted mean ages and metallicities of NSCs and host galaxies.}
    \begin{tabular}{l r r r r r r r r}\hline
       Galaxy  & [M/H]$_\text{NSC}$ & [M/H]$_\text{2\arcsec}$ & [M/H]$_\text{0.5 $R_\text{eff}$}$ & [M/H]$_\text{1.0 $R_\text{eff}$}$ & Age$_\text{NSC}$ & Age$_\text{2\arcsec}$ & Age$_\text{0.5 $R_\text{eff}$}$ & Age$_\text{1.0 $R_\text{eff}$}$ \\
        & (dex) & (dex) & (dex) & (dex) & (Gyr) & (Gyr) & (Gyr) & (Gyr)  \\
       \hline\hline
FCC\,47 & $-$0.11 $\pm$ 0.02 & $-$0.54 $\pm$ 0.02 & $-$0.73 $\pm$ 0.02 & $-$0.78 $\pm$ 0.04 & 13.8 $\pm$ 0.1 & 13.4 $\pm$ 0.2 & 12.8 $\pm$ 0.3 & 12.4 $\pm$ 0.7\\
FCC\,119 & $-$1.02 $\pm$ 0.05 & $-$0.52 $\pm$ 0.03 & $-$0.55 $\pm$ 0.02 & -- & 1.7 $\pm$ 0.5 & 4.5 $\pm$ 0.5 & 6.1 $\pm$ 0.5 & --\\
FCC\,148 & 0.26 $\pm$ 0.03 & $-$0.02 $\pm$ 0.02 & $-$0.29 $\pm$ 0.01 & $-$0.49 $\pm$ 0.03 & 4.3 $\pm$ 0.7 & 3.7 $\pm$ 0.4 & 5.4 $\pm$ 0.4 & 7.2 $\pm$ 0.8\\
FCC\,153 & 0.35 $\pm$ 0.02 & 0.02 $\pm$ 0.02 & $-$0.13 $\pm$ 0.01 & $-$0.21 $\pm$ 0.01 & 9.3 $\pm$ 0.6 & 7.6 $\pm$ 0.4 & 8.9 $\pm$ 0.3 & 9.9 $\pm$ 0.4\\
FCC\,170 & 0.13 $\pm$ 0.03 & $-$0.03 $\pm$ 0.01 & $-$0.21 $\pm$ 0.01 & $-$0.29 $\pm$ 0.01 & 13.0 $\pm$ 0.7 & 14.0 $\pm$ 0.0 & 14.0 $\pm$ 0.0 & 13.9 $\pm$ 0.1\\
FCC\,177 & 0.40 $\pm$ 0.01 & $-$0.05 $\pm$ 0.03 & $-$0.36 $\pm$ 0.02 & $-$0.40 $\pm$ 0.02 & 6.7 $\pm$ 0.3 & 4.4 $\pm$ 0.5 & 7.9 $\pm$ 0.4 & 8.2 $\pm$ 0.7\\
FCC\,182 & $-$0.22 $\pm$ 0.05 & $-$0.25 $\pm$ 0.01 & $-$0.41 $\pm$ 0.01 & $-$0.61 $\pm$ 0.02 & 11.3 $\pm$ 0.9 & 10.8 $\pm$ 0.5 & 10.5 $\pm$ 0.4 & 10.3 $\pm$ 0.6\\
FCC\,188 & $-$1.07 $\pm$ 0.04 & $-$0.76 $\pm$ 0.02 & $-$0.81 $\pm$ 0.02 & -- & 8.5 $\pm$ 0.8 & 10.3 $\pm$ 0.8 & 9.2 $\pm$ 0.8 & --\\
FCC\,190 & 0.02 $\pm$ 0.10 & $-$0.24 $\pm$ 0.01 & $-$0.37 $\pm$ 0.01 & $-$0.49 $\pm$ 0.03 & 7.7 $\pm$ 0.9 & 11.5 $\pm$ 0.5 & 11.9 $\pm$ 0.5 & 13.0 $\pm$ 0.6\\
FCC\,193 & 0.40 $\pm$ 0.00 & 0.04 $\pm$ 0.02 & $-$0.26 $\pm$ 0.01 & $-$0.34 $\pm$ 0.02 & 14.0 $\pm$ 0.0 & 11.0 $\pm$ 0.3 & 10.9 $\pm$ 0.3 & 11.6 $\pm$ 0.5\\
FCC\,202 & $-$0.95 $\pm$ 0.05 & $-$0.39 $\pm$ 0.02 & $-$0.47 $\pm$ 0.02 & $-$0.57 $\pm$ 0.01 & 7.4 $\pm$ 1.1 & 7.3 $\pm$ 0.8 & 7.7 $\pm$ 0.7 & 8.0 $\pm$ 0.8\\
FCC\,211 & $-$1.01 $\pm$ 0.04 & $-$1.00 $\pm$ 0.02 & $-$1.04 $\pm$ 0.02 & $-$1.16 $\pm$ 0.02 & 8.8 $\pm$ 0.8 & 9.5 $\pm$ 0.6 & 9.7 $\pm$ 0.4 & 9.6 $\pm$ 0.5\\
FCC\,215 & $-$1.64 $\pm$ 0.12 & $-$1.05 $\pm$ 0.25 & $-$1.59 $\pm$ 0.15 & $-$2.16 $\pm$ 0.11 & 7.8 $\pm$ 2.4 & 5.3 $\pm$ 3.1 & 9.2 $\pm$ 3.2 & 8.3 $\pm$ 3.5\\
FCC\,222 & $-$1.22 $\pm$ 0.06 & $-$0.60 $\pm$ 0.02 & $-$0.76 $\pm$ 0.03 & $-$0.91 $\pm$ 0.05 & 8.9 $\pm$ 0.8 & 8.1 $\pm$ 0.7 & 8.7 $\pm$ 1.0 & 9.4 $\pm$ 1.2\\
FCC\,223 & $-$1.89 $\pm$ 0.08 & $-$1.13 $\pm$ 0.04 & $-$1.18 $\pm$ 0.04 & -- & 11.7 $\pm$ 1.1 & 8.7 $\pm$ 0.9 & 9.9 $\pm$ 1.1 & --\\
FCC\,227 & $-$1.72 $\pm$ 0.15 & $-$1.73 $\pm$ 0.21 & $-$1.88 $\pm$ 0.19 & -- & 12.4 $\pm$ 1.6 & 8.3 $\pm$ 3.5 & 11.6 $\pm$ 2.4 & --\\
FCC\,245 & $-$0.77 $\pm$ 0.14 & $-$0.62 $\pm$ 0.06 & $-$0.83 $\pm$ 0.04 & $-$1.14 $\pm$ 0.07 & 8.2 $\pm$ 2.1 & 8.4 $\pm$ 1.3 & 10.3 $\pm$ 1.0 & 12.2 $\pm$ 1.7\\
FCC\,249 & 0.21 $\pm$ 0.02 & $-$0.28 $\pm$ 0.02 & $-$0.50 $\pm$ 0.01 & $-$0.66 $\pm$ 0.02 & 13.7 $\pm$ 0.3 & 13.7 $\pm$ 0.1 & 12.7 $\pm$ 0.1 & 12.1 $\pm$ 0.2\\
FCC\,255 & $-$0.16 $\pm$ 0.04 & $-$0.28 $\pm$ 0.02 & $-$0.31 $\pm$ 0.02 & $-$0.43 $\pm$ 0.02 & 3.7 $\pm$ 0.7 & 5.1 $\pm$ 0.6 & 5.1 $\pm$ 0.5 & 6.0 $\pm$ 0.6\\
FCC\,277 & 0.16 $\pm$ 0.04 & $-$0.31 $\pm$ 0.01 & $-$0.46 $\pm$ 0.01 & $-$0.57 $\pm$ 0.01 & 9.7 $\pm$ 0.8 & 9.7 $\pm$ 0.4 & 10.6 $\pm$ 0.3 & 11.1 $\pm$ 0.5\\
FCC\,301 & $-$0.23 $\pm$ 0.05 & $-$0.24 $\pm$ 0.02 & $-$0.53 $\pm$ 0.01 & $-$0.74 $\pm$ 0.02 & 6.0 $\pm$ 1.1 & 6.2 $\pm$ 0.6 & 7.8 $\pm$ 0.6 & 9.2 $\pm$ 0.7\\
FCC\,310 & 0.27 $\pm$ 0.04 & $-$0.33 $\pm$ 0.02 & $-$0.50 $\pm$ 0.04 & $-$0.55 $\pm$ 0.08 & 7.5 $\pm$ 0.8 & 9.6 $\pm$ 0.5 & 10.3 $\pm$ 1.2 & 10.3 $\pm$ 1.8\\
FCCB\,1241 & $-$0.88 $\pm$ 0.13 & $-$1.33 $\pm$ 0.05 & $-$1.21 $\pm$ 0.05 & $-$1.43 $\pm$ 0.07 & 10.6 $\pm$ 2.1 & 11.8 $\pm$ 1.1 & 12.0 $\pm$ 1.3 & 12.2 $\pm$ 1.3\\
VCC\,990 & $-$0.20 $\pm$ 0.04 & $-$0.50 $\pm$ 0.02 & $-$0.55 $\pm$ 0.02 & $-$0.65 $\pm$ 0.02 & 4.4 $\pm$ 0.9 & 5.8 $\pm$ 0.6 & 6.0 $\pm$ 0.6 & 6.7 $\pm$ 0.6\\
VCC\,2019 & 0.06 $\pm$ 0.05 & $-$0.49 $\pm$ 0.02 & $-$0.68 $\pm$ 0.02 & $-$0.81 $\pm$ 0.03 & 3.2 $\pm$ 0.7 & 5.9 $\pm$ 0.7 & 7.2 $\pm$ 0.7 & 9.3 $\pm$ 0.9\\
\hline
    \end{tabular}
     \label{tab:results}
\end{table*}

\subsection{Star formation histories}
\label{sect:SFHs_MDFs}
By applying regularisation during the pPXF fit, smooth distributions of ages and metallicities that best fit the spectra are recovered. Their weights can be used to infer normalised SFHs, as shown in Fig.  \ref{fig:SFHs_coloured}. Each panel shows the SFH of the NSC (top), the galaxy at 2\arcsec\,from the centre (middle), and the host galaxy at 1.0 or 0.5 $R_\text{eff}$ (bottom). The SFHs are normalised such that the peak of the distribution is at 1 for visualisation of the shape and the distributions are colour-coded to show the mean metallicity in each age bin. We only show the galaxies where the NSC and galaxy spectra have sufficient $S/N$ to ensure a reasonable application of the regularisation approach. For this reason, FCC\,215 and FCC\,227 are not shown.

We find a large diversity in SFH shapes for both the NSCs and their host galaxies. Generally, the SFHs at 2\arcsec and at 1.0 $R_\text{eff}$ have similar shapes, while the NSCs often show a very different behaviour from the host galaxy. At lowest galaxy masses ($M_\text{gal} < 10^9 M_\sun$, top row), the host galaxies predominantly show single peaks in the SFH with mean ages around 8 Gyr, but the SFHs of their NSCs show a greater variety, also displaying very old stellar ages and low metallicities. These could be explained by early star formation in the NSC, or by the accretion of GCs which are typically old and metal-poor in such dwarf galaxies (e.g. the GCs of the Fornax dwarf spheroidal galaxy \citealt{Larsen2012}). 
FCC\,119 stands out from this with very young ages in the NSC and the presence of ionised gas indicative of star formation.

The NSCs in the most massive galaxies of our sample (bottom row) are dominated by a single peak at old ages with high metallicities, while the host galaxies exhibit similar SFHs but offset at lower metallicities.
At intermediate host masses, the diversity in NSC SFHs increases. For galaxies with masses around $10^9 M_\sun$ (second and third row), we find several cases in which the NSC SFH is extended or shows multiple populations with different ages and metallicities, including young stellar populations.

\subsection{Metallicity distributions}
The metallicity distributions for the NSCs and their host galaxies (at 2\arcsec\,and 1 $R_\text{eff}$ where available) are shown in Fig. \ref{fig:MDFs_coloured}, coloured by the mean age per metallicity bin. 
The distributions are less smooth than the SFHs in Fig. \ref{fig:SFHs_coloured} because the E-MILES SSP models give only 12 different metallicities. In general, a trend with mass is seen in the sense that the mean metallicity increases with mass of both the host and NSC, as will be discussed in more detail in Sect. \ref{sect:MZR}. 

In the GC-accretion scenario of NSC formation, the formed NSC should reflect the metallicity distribution of the accreted GCs. Although these GCs (or younger star clusters) might have been different from today's GC population, we also show the metallicity distributions of GCs in Fig. \ref{fig:MDFs_coloured} for a qualitative comparison.
Those are obtained for all galaxies that have photometric GC catalogues from the ACSFCS or ACSVCS available \citep{Jordan2007, Jordan2015}. We only use GCs with projected distances to the galaxy centres < 1.0 $R_\text{eff}$.
The catalogues provide accurate ($g - z$) colours that we translated to metallicities using the colour-metallicity relation from \cite{Fahrion2020c}, which was derived from the F3D GC sample. For visualisation purposes, we normalised the GC metallicity distributions to match the peak height of the galaxy metallicity mass fractions. Unfortunately, most of the dwarf galaxies have no photometric GC catalogues available.

For low-mass galaxies, we find significant mass fractions at low metallicities, while the NSCs of the high-mass galaxies are dominated by high metallicities. 
Comparing the GC and galaxy metallicity distributions shows a good match at high metallicities, but most galaxy metallicity distributions do not extend to low metallicities and are on average more metal-rich than the GCs. This is not surprising as most massive galaxies show two populations of GCs that differ in their metallicity with a metal-rich population that was formed \textit{in-situ} in the host, and a metal-poor population stemming from accreted dwarf galaxies (e.g.  \citealt{Harris1991, Peng2006, Lamers2017, Beasley2020, Fahrion2020b, Fahrion2020c}). 

We find several cases where the NSC is significantly more metal-rich than the most metal-rich GCs observed today, for example in FCC\,170, FCC\,193, and FCC\,249. This could indicate that these NSCs were either formed from \textit{in-situ} star formation, or out of a population of massive, metal-rich star clusters that are no longer present in the central regions of these galaxies.
In other cases, the GC metallicities could naturally explain the metallicities found in the NSC, for example in FCC\,202 and FCC\,255. In these cases, the NSCs are significantly younger (< 7 Gyr), than typical GCs (> 8 Gyr \citealt{Puzia2005}), which then would require additional \textit{in-situ} star formation or accretion of young massive star clusters. Unfortunately, we do not have reliable age measurements of a large sample of GCs from these galaxies. The dwarf galaxies are the most likely hosts of NSCs formed through GCs due to their low metallicities and old ages.

\subsection{Mass-metallicity relation}
\label{sect:MZR}

\begin{figure*}
    \centering
    \includegraphics[width=0.95\textwidth]{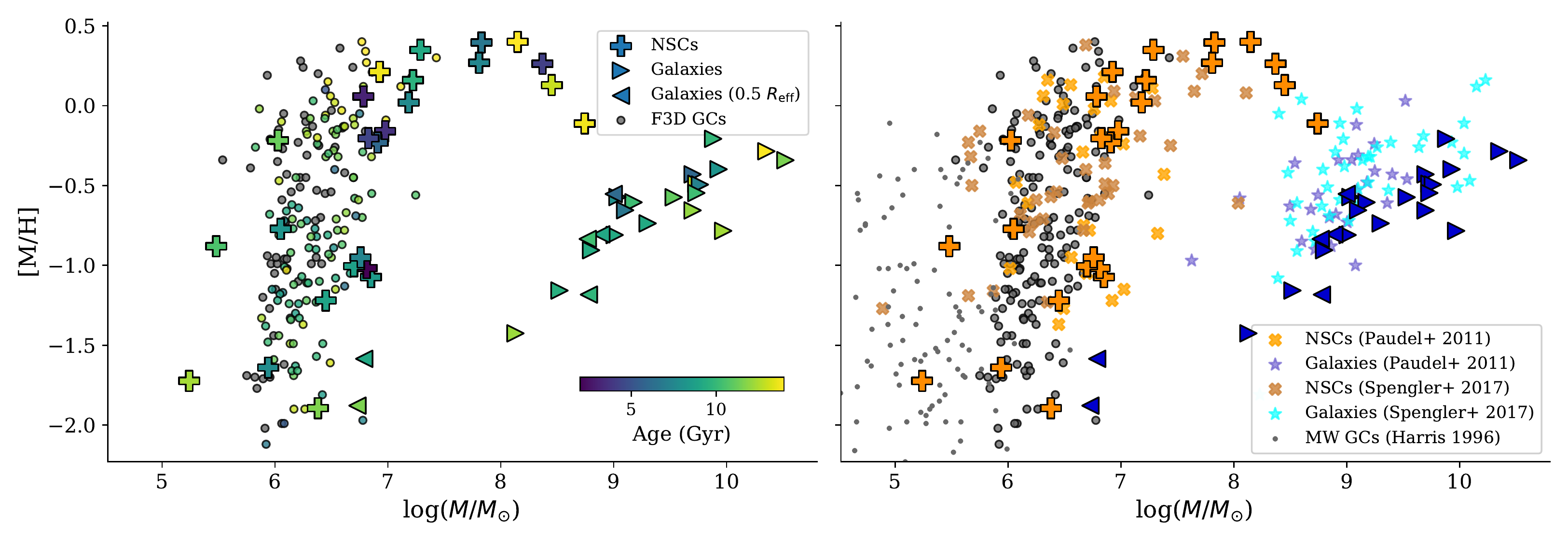}
    \caption{Mass-metallicity plane. Both panels show mass and metallicity for the host galaxies of our sample at 1.0 $R_\text{eff}$ where available (rightward triangles, otherwise leftward triangles), NSCs (plus symbols) and GCs (coloured circles, from \citealt{Fahrion2020b}).
    \textit{Left}: The different components are colour-coded by the mean age, inferred from full spectrum fitting.
    \textit{Right}: For comparison, we include literature NSCs (orange and brown crosses) and their hosts galaxies (bright blue and purple stars) from \cite{Paudel2011} and \cite{Spengler2017}, and MW GCs (small grey dots) from \cite{Harris1996}.}
    \label{fig:MZR_full}
\end{figure*}

We show the mass-metallicity plane of the NSCs and the host galaxies (at 1.0 $R_\text{eff}$ where available) in Fig. \ref{fig:MZR_full}. For comparison, we added the F3D GCs \citep{Fahrion2020b, Fahrion2020c} which are clearly separated from the host galaxies in the mass-metallicity plane. While the galaxies follow a clear mass-metallicity relation (see e.g. \citealt{Gallazzi2005, Kirby2013}), the GCs only show a large scatter in both parameters (see also \citealt{Zhang2018, Fahrion2020c}). The NSCs, however, show a different behaviour. For $M_\text{NSC} < 10^7 M_\sun$, the NSCs lie in the same region as GCs, covering a large spread in metallicities. Higher mass NSCs ($M_\text{NSC} > 10^7 M_\sun$) are exclusively metal-rich, and exceed the metallicity of galaxies with similar masses by 1.5 dex. 
Interestingly, there are a few high mass GCs ($M_\text{GC} \sim 10^7 M_\sun$) that also seem to follow the trend of these massive NSCs. Those could be candidates of ultra-compact dwarf galaxies (UCDs, \citealt{Hilker1999, Drinkwater2000}), which are compact stellar systems that exceed the typical masses and sizes of GCs and were suggested to be the remnant NSCs of disrupted dwarf galaxies \citep{Strader2013}.



The left panel of Fig. \ref{fig:MZR_full} shows the different components colour-coded by their ages. In the massive NSCs ($M_\text{NSC} \gtrsim 10^7 M_\sun$), we find either extremely old ($>$ 10 Gyr) or young ages ($<$ 5 Gyr), while the low-mass NSCs ($M_\text{NSC} < 5 \times 10^6 M_\sun$) are generally old (8 - 10 Gyr), comparable to the ages of GCs. All galaxies have mean ages $>$ 5 Gyr at 1.0 $R_\text{eff}$ as is expected since our sample consists of early-type galaxies which have ceased star formation globally. Nonetheless, those can still show star formation in their centres (e.g. \citealt{Richtler2020}).

The left panel in Fig. \ref{fig:MZR_full} shows a comparison of our sample to literature values. We show NSCs and their hosts from \cite{Paudel2011} and \cite{Spengler2017}, and MW GCs from (\citealt{Harris1996}, catalogue version 2010). The MW GCs extend our F3D GC sample to lower masses that are not accessible with the current data due to the intrinsic brightness of these low mass GCs at distances of the Fornax galaxy cluster. Despite the different mass ranges, they span roughly the same range in metallicity, although even metal-rich bulge GCs only reach [Fe/H] $\sim -0.5 - -0.15$ dex \citep{Munoz2017, Munoz2018, Munoz2020} and thus very metal-rich GCs are lacking in the MW.

The literature NSCs cover comparable ranges in masses and metallicities as our sample. They show a similar trend with increasing NSC mass as our sample: there are only a few metal-poor NSCs with masses $> 10^7 M_\sun$, while the low-mass NSCs cover the mass-metallicity space of the GCs. The literature galaxy sample spans roughly the same ranges of masses and metallicities as the galaxies studied here, but our sample extends to lower galaxy masses, which might reflect in the lack of very metal-poor NSCs in the literature sample.


\subsection{Constraining NSC formation from stellar populations}
\label{sect:pops_different_channels}
In this section we take the results presented above together to draw conclusions for individual galaxies, grouped by similar NSC properties. We focus on the radial metallicity profiles (Fig. \ref{fig:radial_profiles}), the SFHs and metallicity distributions (Figs. \ref{fig:SFHs_coloured}, \ref{fig:MDFs_coloured}). The results are summarised in Fig. \ref{fig:dominant_channel}. With our stellar population analysis, we can compare the ages and metallicities of the unresolved NSCs to the host galaxies and the observed GCs. While we can identify accretion signatures of old, metal-poor GCs, for young and metal-rich populations we cannot discern between \textit{in-situ} star formation happening directly in the NSC or in central star clusters that were rapidly accreted into the NSC.

\subsubsection{Most massive galaxies}
The most massive galaxies in our sample ($M_\text{gal} > 10^{10} M_\sun$), FCC\,47, FCC\,170, and FCC\,193, also host the most massive NSCs ($> 10^8 M_\sun$). Their NSCs are significantly more metal-rich than the hosts and any GCs, and are older than the host galaxies. This could be explained by an inside-out formation scenario, where the central regions of the galaxies assembled first and were rapidly enriched due to efficient star formation, while enrichment in the outskirts was slower and less efficient. In \cite{Fahrion2019b}, we argued that the old age and high metallicity of the NSC in FCC\,47 indicates an efficient formation at early times. Similar arguments were made in the detailed analysis of FCC\,170 by \cite{Pinna2019a} using the same F3D MUSE data which showed that FCC\,170 assembled very rapidly and efficiently at early times (see also \citealt{Poci2021}). Their SFH of the central region of FCC\,170 is similar to the one presented here. Alternatively, the high metallicity could be a result of the infall and merger of a population of young and metal-rich massive star clusters.

The high NSC masses also indicate formation from efficient central star formation rather than from ancient GCs as their high masses ($M_\text{NSC} > 10^8 M_\sun$) would require the accretion of hundreds of metal-rich star clusters with masses $\gtrsim 10^6 M_\sun$ - already close to the cut-off mass typically observed for young massive star clusters in highly interacting galaxies \citep{Adamo2020}.
An \textit{in-situ} origin is further supported by the significant rotation in the NSCs of FCC\,170 and FCC\,47 found based on high angular resolution integral-field spectroscopy with VLT/SINFONI \citep{Lyubenova2019}. \cite{Fahrion2019b} showed that the NSC of FCC\,47 is a kinematically decoupled component, which likely indicates a major merger that has shaped the kinematic properties of this NSC. Such a merger could also explain the slightly younger, more metal-poor populations we found in the outskirts of FCC\,47 and FCC\,193, which could be accreted material from a lower mass galaxy.

\subsubsection{Lower-mass edge-on S0 galaxies}
FCC\,153 and FCC\,177 are edge-on S0 galaxies like FCC\,170, but at significantly lower galaxy mass (see e.g. \citealt{Pinna2019b, Poci2021} for a detailed analysis using the F3D MUSE data). \cite{Pinna2019b} and the Voronoi binned radial profiles presented in this work generally find young ages toward the galaxy centres. However, our background-subtracted NSC spectra find the NSCs to be older than their immediate surrounding although at a similar high metallicity, but with a minor young stellar population. The NSCs are more metal-rich than any GCs found in the galaxies. The high NSC metallicity and the presence of young stars indicate a large contribution from \textit{in-situ} star formation to their build-up. This star formation might have happened directly in the NSCs or in clustered star formation in the central regions of these galaxies.

\subsubsection{Galaxies with young central populations}
FCC\,119 and FCC\,148 are the only two sample galaxies that show the presence of ionised gas in their NSC as traced by the presence of H$\alpha$, H$\beta$, [NII], [SII] and [OIII] lines in the MUSE spectra (see also the emission line maps in \citealt{Iodice2019}). Our analysis uncovered young stellar populations ($< 3$ Gyr) in both NSCs, while their host stellar bodies are older. The same was found for VCC\,2019 although without the clear presence of ionised gas. This ongoing or recent star formation is a clear sign of \textit{in-situ} star formation building up the NSCs. However, FCC\,119 also shows an older, more metal-poor component in its NSC and is on average more metal-poor than the host galaxy. This drop in metallicity could indicate additional accretion of metal-poor GCs from the galaxy outskirts to the centre.

FCC\,301 and VCC\,990 have similar galaxy ($M_\text{gal} \sim 10^{9} M_\sun$) and NSC masses ($M_\text{NSC} \sim 7 \times 10^6 M_\sun$), and their stellar population properties are also similar to that of FCC\,255 which is slightly more massive. In these three galaxies, the NSCs are younger and more metal-rich than their hosts, their SFHs show a single peak at age $\sim$ 3 Gyr. The young ages and high metallicities indicate NSC formation dominated by \textit{in-situ} star formation either in recently accreted young massive star clusters or directly in the NSC.

\subsubsection{Galaxies with old, metal-rich NSCs}
FCC\,190, FCC\,310, FCC\,277, and FCC\,249 have similar stellar masses ($\sim 3 - 5 \times 10^9 M_\sun$). Their NSCs are more metal-rich than the hosts and any GCs. The NSCs and galaxies are characterised by old stellar populations ($>$ 10 Gyr) and FCC\,190 and FCC\,277 show a minor young, metal-rich population in its NSC. The high metallicities of the NSCs as well as the extended SFHs in FCC\,190, FCC\,310, and FCC\,277 are in agreement with formation from multiple episodes of \textit{in-situ} star formation or the repeated accretion of young star clusters.

Based on the presence of a kinematically cold component found through Schwarzschild dynamical modelling of high spatial resolution VLT/SINFONI data of FCC\,277, \cite{Lyubenova2013} came to conclusion that gas infall likely played an important role in the formation of the NSC, hence further supporting our conclusions from the stellar population properties. However, they also identified a minor population with counter-rotating orbits that indicate addition contribution from a compact object that has merged with the NSC. Consequently, this galaxy is an example where the stellar population properties indicate that the formation of the NSC was dominated by \textit{in-situ} star formation, but additional GC merging (or merging of young star clusters) is required to also explain the complex kinematics of this NSC. This illustrates, that while stellar population analysis is crucial to identify the major NSC formation scenario, detailed dynamical modelling is required to fully unveil the formation history of a NSC.

\subsubsection{Galaxies with evidence of both NSC formation scenarios}
The radial metallicity profiles of FCC\,188, FCC\,202, FCC\,222, and FCC\,245 all clearly show a drop in metallicity in the galaxy centres and their NSCs are on average more metal-poor than the host galaxies. This drop in metallicity is indicating formation from the infall of metal-poor GCs, but the SFHs of these galaxies all show a secondary component with a higher metallicity and significantly younger age (also seen in FCC\,119). This metal-rich, young population can be explained by additional \textit{in-situ} star formation or the accretion of a young star cluster. The NSCs of these galaxies are therefore examples of NSCs were both formation channels are acting together.

FCC\,182 is an especially intriguing case. The NSC has roughly the same metallicity as the galaxy at 2\arcsec, but the radial metallicity profile of FCC\,182 reveals that the NSC still constitutes a dip in the metallicity profile. 
This suggests formation from in-spiral of one or two metal-poor GCs to the centre which could also explain the low NSC mass ($M_\text{NSC} \sim 10^6 M_\sun$). In contrast, the GC population of FCC\,182 today does not exhibit GCs that reach the metallicity of the NSC and the extended NSC SFH would rather indicate formation from prolonged \textit{in-situ} star formation or the accretion of multiple star clusters of different ages. However, also the SFH shows a minor old, metal-poor component in addition to higher metallicities and younger ages. To fully explain the radial metallicity profile and the complex SFH, a likely contribution from both NSC formation channels is required.

\subsubsection{Dwarf galaxies with old, metal-poor NSCs}
FCC\,215, FCC\,211, and FCC\,223 are dwarf galaxies, in which the radial metallicity profiles reveal a drop in metallicity in the centre and the NSCs are significantly more metal-poor than their hosts. These three galaxies are prime candidates for NSC formation dominated by the accretion of old and metal-poor GCs (see also \citealt{Fahrion2020a}).

FCC\,B1241 and FCC\,227 have the least massive NSCs in our sample with stellar masses $< 10^6 M_\sun$. We found that the NSC of FCCB\,1241 is on average more metal-rich than the host galaxy, but also shows an old, metal-poor component, indicating a mixture of both NSC formation channels acting in its growth. The $S/N$ in the NSC and galaxy spectra of FCC\,227 do not allow to determine a reliable metallicity profile, but the mass, age, and metallicity of the NSCs are fully in agreement with typical GCs.

\subsection{Trend with galaxy and nuclear star cluster masses}
\begin{figure}
    \centering
    \includegraphics[width=0.47\textwidth]{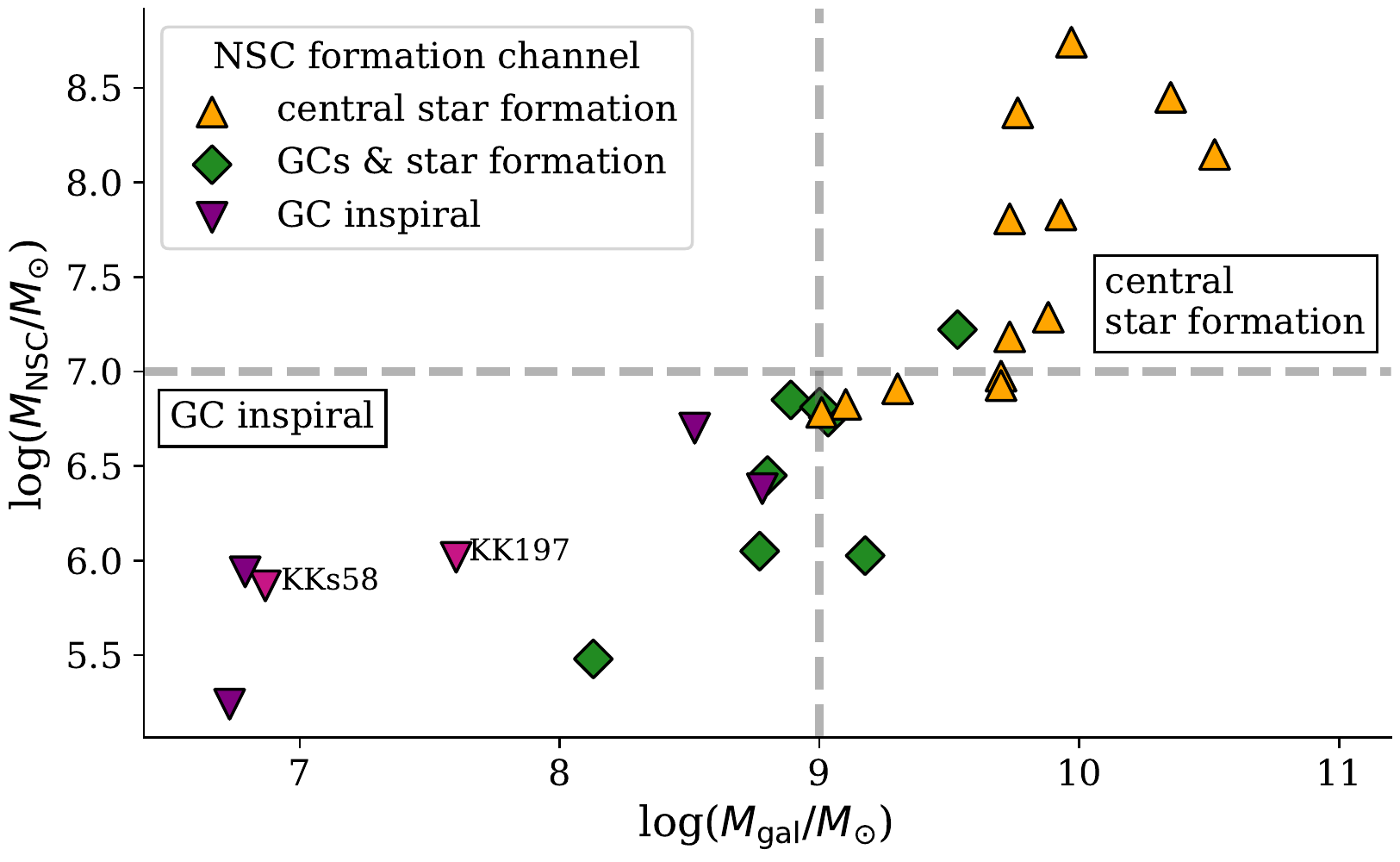}
    \caption{NSC mass versus galaxy mass for our sample. The different colours and symbols refer to the identified dominant NSC formation channel. We differentiate between dominated by central \textit{in-situ} star formation either directly at the NSC or in central star clusters that were rapidly accreted as determined from high metallicities or young ages (orange), formation through GC-accretion as evident from central declines in metallicity (purple), and galaxies in which both channels are required to explain the NSC properties (green). We included KKs\,58 and KK\,197, two nucleated dwarf galaxies with NSCs formed through GC-accretion \citep{Fahrion2020a}. The dashed lines indicate the transition NSC and galaxy masses where the dominant channel changes.}
    \label{fig:dominant_channel}
\end{figure}

Figure \ref{fig:dominant_channel} summarises the discussion in the previous section. It shows the NSC-galaxy mass distribution of our sample, and the different symbols indicate the identified dominant NSC formation channel. We group the galaxies into three different categories based on the identified dominant NSC formation channel: formation from the inspiral of old, metal-poor GCs as evident from metal-poor populations in the NSC, formation as a result of central star formation as well as a mixed scenario that invokes both channels. We note that central star formation can refer to both the classical \textit{in-situ} formation channel, where the star formation results from accreted gas at the NSC position, as well as to clustered star formation in the central regions of the galaxies.
 The two nucleated dwarf galaxies KK\,197 and KKs\,58 that were analysed in \cite{Fahrion2020a} are also included. Based on the low metallicity of their NSCs in comparison to their hosts their NSCs were likely formed via the in-spiral of a few GCs to the centre. 

Figure \ref{fig:dominant_channel} reveals a trend in dominant NSC formation channel both with galaxy and NSC mass. At the lowest galaxy and NSC masses, we find NSCs predominantly form via the infall of GCs, while at the highest galaxy and NSC masses NSC formation is dominated by central star formation (either in young star clusters close to the centre or in the NSC directly). At intermediate masses ($M_\text{gal} \sim 10^9 M_\sun$, $M_\text{NSC} \sim 5 \times 10^6$), we find galaxies in which either formation channel is responsible for the NSC properties observed today, as well as several cases in which both channels are acting together.

\section{Discussion}
\label{sect:discussion}
In the following we discuss our results by comparing them to literature values and other studies. 

\subsection{Comparison with literature values}
\cite{Iodice2019a} presented luminosity-weighted stellar population properties from line-strength measurements for the F3D galaxies. Their Table 3 lists the mean age, metallicity, and light-element abundance ratio within the central region ($R < 0.5\,R_\text{eff}$), in a similar region as our spectra extracted at 0.5 $R_\text{eff}$. We find generally a good match between this work and the results from \cite{Iodice2019a} despite the different spatial extraction regions and measurement methods. 

The Fornax dwarf galaxies in our sample were previously analysed in \cite{Johnston2020}. That work uses \textsc{buddi} \citep{Johnston2017}, a tool to decompose a MUSE datacube into different spatial components describing the NSC and the host galaxy, similar to our approach. While we used \textsc{imfit} to build a model only to estimate the contribution of the underlying galaxy light to the NSC spectrum, \cite{Johnston2020} used their modelling technique to derive model spectra of every component by considering the MUSE cubes as a series of narrow-band images. However, because each component is described by a predefined function, for example a S\'ersic profile, there is only one model spectrum per component and thus internal radial stellar population gradients cannot be represented. 

\begin{figure}
    \centering
    \includegraphics[width=0.43\textwidth]{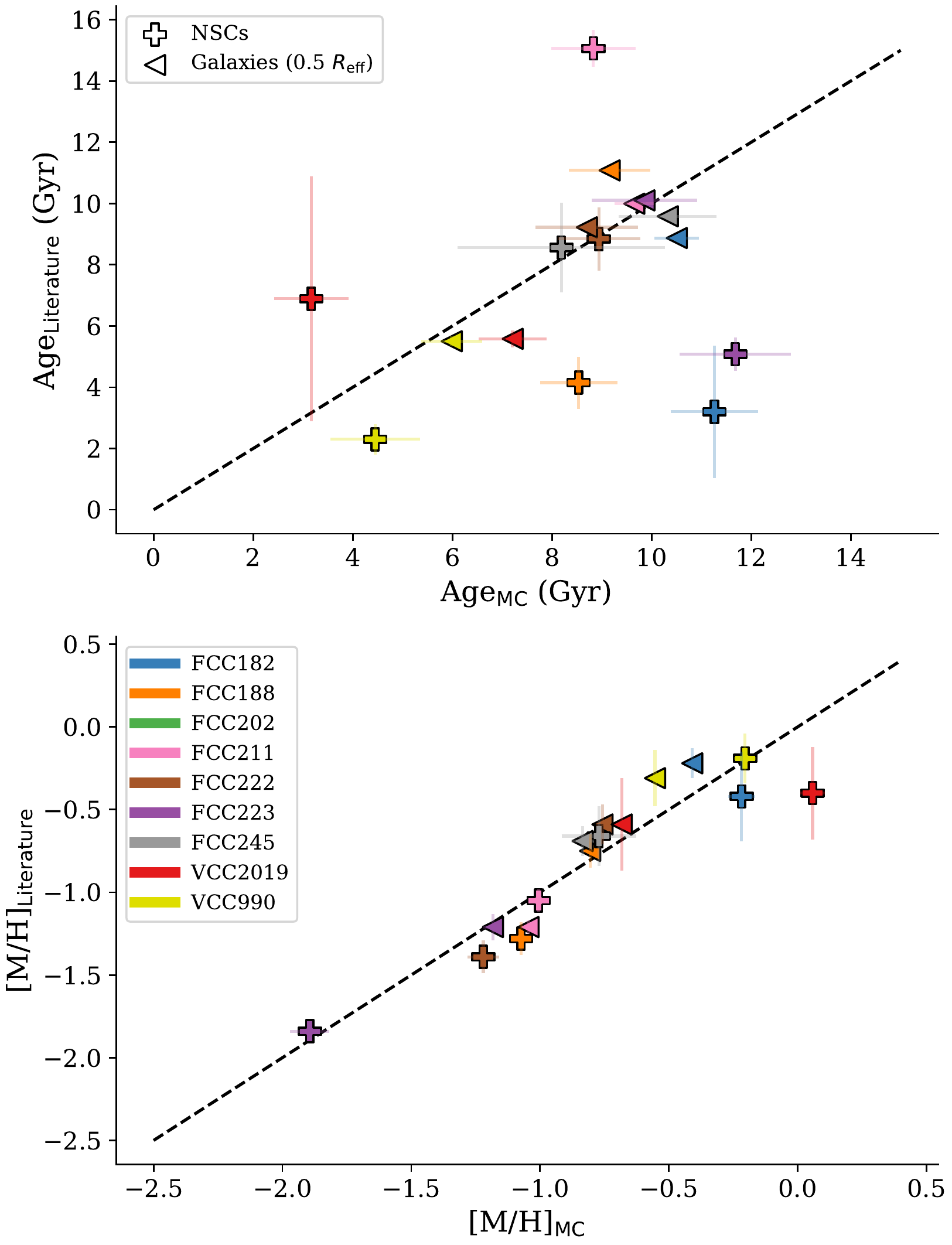}
    \caption{Comparison between the NSC (crosses) and host galaxy (triangles, at 0.5 $R_\text{eff}$) age (\textit{top}) and metallicity (\textit{bottom}) from the literature and this work. The different colours refer to the different galaxies. VCC\,990 was analysed by \cite{Paudel2011}, VCC\,2019 by \cite{Spengler2017}, and the other galaxies by \cite{Johnston2020}. The dashed lines refer to the one-to-one relation.}
    \label{fig:johnston_comp}
\end{figure}

In Fig. \ref{fig:johnston_comp}, we plot the NSC and galaxy ages and metallicities for the galaxies that were studied both here and by \cite{Johnston2020} with full spectral fitting. We also add the Virgo galaxies VCC\,990, which was analysed with spectroscopy and Lick line-strength indices in \cite{Paudel2011}, and VCC\,2019, which was among the galaxies studied with broad-band photometry by \cite{Spengler2017}. 
The metallicities agree well between our work and the literature. The ages show a larger scatter, but in general the overall trend (young versus old) is recovered. The only exceptions are the NSCs of FCC\,182 and FCC\,223, which we find to be significantly older than estimated by \cite{Johnston2020}.  The reason for this discrepancy is unknown, but could be caused by the different approaches to extract the stellar population properties. \cite{Johnston2020} also used \textsc{pPXF} with regularisation and the MILES SSP model library, but with a different grid of models based on the so-called Padova isochrones \citep{Girardi2000}. The scatter in ages might also reflect the well-known difficulty of deriving accurate stellar ages from integrated spectroscopy (e.g. \citealt{Usher2019}). Further, the wavelength coverage of MUSE instrument is lacking many age-sensitive absorption features. We note that the line-strength measurements presented in \cite{Iodice2019a} also found an old age for FCC\,182 of 12.6 Gyr in the central region of the galaxy. 

\subsection{Complexity in nuclear star clusters}
NSCs are known to be complex stellar systems that can show a large diversity of characteristics. Some NSCs show non-spherical density distributions with flattened structures \citep{Boeker2002}, and many show multiple stellar populations \citep{Walcher2006, Lyubenova2013, Kacharov2018}, often with significant contributions from young stellar populations \citep{Rossa2006, Paudel2011}. This complexity is also often seen in the internal kinematics of NSCs as some show significant rotation and distinct dynamical structures \citep{Seth2008, Seth2010, Lyubenova2013, Lyubenova2019}. 

In this work, we focused on the global stellar population properties of NSCs and found a large variety in ages, metallicities, SFHs, and metallicity distributions. Many of the studied NSCs show extended SFHs or multiple populations as also observed in other studies. For example, \cite{Kacharov2018} studied the NSC of six nearby galaxies ($M_\text{gal} \sim 2 - 8 \times 10^9 M_\sun$) with high resolution spectroscopy. They found prolonged NSC SFHs which include young stars.
This is in line with several other studies that have found NSCs of dwarf ellipticals ($M_\text{gal} \gtrsim 10^8 M_\sun$) to be generally younger than their host galaxy \citep{Paudel2011, Guerou2015, Paudel2020}. In our sample, we found NSCs of dwarf galaxies both older and younger than the host galaxies, but many also share the age of their host. We note that our sample extends to lower galaxy masses than \cite{Paudel2011} for example.

At the lowest galaxy and NSC masses studied here, we found that the NSCs share many similarities with GCs in their low masses, mean low metallicities, and old ages. This is a natural consequence if those NSCs were formed from GCs, but has implications for identifying remnant NSCs of disrupted satellite galaxies that might be hiding in the rich GC populations of massive galaxies \citep{Pfeffer2014, Pfeffer2016}, or even in the GC population of the MW (e.g. \citealt{Sakari2020, Pfeffer2021}). Such an origin was proposed, for example, for $\omega$Centauri, the most massive GC in the MW \citep{Lee1999, HilkerRichtner2000, BekkiFreeman2003, vandeVen2006, Villanova2014}\, although the most compelling case has been made for M\,54, which is located at the centre of the Sagittarius dwarf galaxy and shows a large spread in iron abundance as well as an extended SFH \citep{Ibata1997, Bellazzini2008, Sills2019, AlfaroCuello2019}. An origin as stripped NSCs was also proposed for UCDs to explain their high masses and large sizes (see e.g. \citealt{Pfeffer2013, Strader2013, Norris2015}), but they could also constitute the high-mass end of the genuine star cluster population \citep{Mieske2002, KisslerPatig2006}. 
In the most massive UCDs, extended SFHs \citep {NorrisKannappan2011} and even SMBHs were found \citep{Seth2014, Ahn2017, Ahn2018, Afanasiev2018}, but the origin of low-mass UCDs is less clear \citep{Fahrion2019a}. Those are often metal-poor and of similar mass as some GCs, but as this work shows, these properties also agree with the properties of low-mass NSCs.

\subsection{Diversity in NSC formation channels}
Based on the shape of the nucleation fraction function with galaxy mass which shows a peak at $M_\text{gal} \sim 10^9 M_\sun$ and results from stellar populations, \cite{Neumayer2020} suggested a transition of the dominant NSC formation channel from GC-accretion to \textit{in-situ} star formation with increasing galaxy mass.

In this work, we confronted this suggestion with observations of galaxies that span a broad range in both NSC and galaxy masses and that are studied in a homogeneous fashion. We found that many of the studied NSCs exhibit complex SFHs and multiple populations of different metallicities. All NSCs residing in galaxies more massive than $10^9 M_\sun$ are significantly more metal-rich than the host galaxy, indicating formation from efficient central star formation directly at the bottom of the galactic potential well or as clustered star formation in the central regions. In contrast, at the lowest galaxy masses, the NSCs appear to be have formed through the accretion of a few metal-poor GCs. Those dwarf galaxies exhibit radial metallicity profiles with clear drops in metallicity at the galaxy centre. At intermediate galaxy and NSC masses, we found several cases of NSC SFHs showing evidence of both channels contributing to the NSC formation. 

Underlying this trend with galaxy mass, we find a transition of the dominant NSC formation channel with NSC mass as evident in Fig. \ref{fig:MZR_full}.
For masses $M_\text{NSC} > 10^7 M_\sun$, the NSCs clearly separate from GCs in their mass, metallicity, and age diversity. The low-mass NSCs in our sample are overall old ($\gtrsim$ 7 Gyr), but span a large range in metallicity - consistent with GCs of similar mass. Conversely, more massive NSCs can also show very young stellar populations. 
Around NSC masses of $\sim 10^7 M_\sun$, the largest diversity in age and metallicity is found. This transition mass $\sim 10^7 M_\sun$ is probably connected to a possible mass limit of GC formation. As \cite{Norris2019} argue, there appears to be an upper mass limit of the ancient star cluster population $\sim 5 \times 10^7 M_\sun$. As star clusters of this mass should be scarce due to their position in the high mass tail of the GC mass function, this maximum mass would also suggest a mass limit for NSCs formed solely from GCs because more massive NSCs would require the merger of hundreds of GCs. 

In summary, as Fig. \ref{fig:dominant_channel} illustrates, we indeed find a transition of the dominant NSC formation channel with galaxy and NSC mass: while the short dynamical friction timescales in low-mass galaxies can form low-mass NSCs from the accretion of a few ancient metal-poor GCs, central \textit{in-situ} star formation or the accretion of young and enriched star clusters is responsible for the mass build-up in the most massive NSCs. At intermediate masses, both channels can contribute.


Such a transition of dominant NSC formation channel is also likely reflected in the galaxy-NSC mass relation. While GC-accretion can build the low-mass NSCs of dwarf galaxies, this process is not efficient enough to reach the high NSC masses found in more massive galaxies. In those, \textit{in-situ} star formation appears to be effective in building massive NSCs, often already at early times. The large diversity in SFHs at high NSC and galaxy masses suggests that NSC formation in these systems is correlated with the evolutionary history of the galaxy, for example due to the availability of gas or the occurrence of major mergers which might prevent GC-inspiral \citep{Leung2020}, but can trigger central star formation \citep{Schoedel2020}.

\section{Conclusion}
\label{sect:conclusion}
In this paper, we presented a stellar population analysis of 25 nucleated early-type and dwarf galaxies mostly in the Fornax cluster to derive the most likely NSC formation path. We extracted the mean ages and metallicities as well as SFHs and metallicity distributions and analysed those properties with focus on the NSC formation. Our main results are:
\begin{itemize}
    \item We found a large diversity of NSC SFHs. The NSCs of massive galaxies are metal-rich, and we found both very old NSCs and NSCs with ongoing or recent \textit{in-situ} star formation. At intermediate galaxy masses ($M_\text{gal} \sim 10^9 M_\sun$), the diversity in SFHs even increases. We found cases of single-peaked SFHs and SFHs with multiple bursts. Towards the lowest galaxy masses ($M_\text{gal} < 10^9 M_\sun$), the NSCs tend to be as old as their host galaxies.
    \item Radial metallicity profiles reveal that the NSCs of massive galaxies ($M_\text{gal} > 10^9 M_\sun$) are more metal-rich than their host galaxy. In contrast, at lower masses, several of the studied galaxies show a clear drop in metallicity at the location of the NSC.
    \item Putting NSCs, GCs, and galaxies on a common mass-metallicity plane, we found that the population of massive NSCs ($> 10^{7} M_\sun$) is distinct from the less massive NSCs, which occupy the same region in the mass-metallicity plane as GCs. The massive NSCs, however, branch off in mass at exclusively high metallicities and can show either extremely old or young populations. 
    \item Our results suggest different formation paths of high and low-mass NSCs to explain their different properties. The high masses and high metallicities of the massive NSCs found in massive galaxies suggest formation from \textit{in-situ} star formation or the accretion of young enriched star clusters. In contrast, the low NSC metallicities found in the low-mass galaxies of our sample are likely the result of infalling metal-poor GCs forming these NSCs. Furthermore, we find several galaxies where both channels are required to explain the properties of their NSCs.
    \item We found clear evidence for a shift of the dominant NSC formation channel from GC accretion at low NSC and galaxy masses towards central \textit{in-situ} star formation forming the most massive NSCs in our sample. The transition appears to happen for NSC masses with $M_\text{NSC} \sim 10^7 M_\sun$ and galaxies with masses $M_\text{gal} \sim 10^9 M_\sun$.
\end{itemize}
Our work shows that NSCs are diverse objects. Some appear as simple stellar systems that are indistinguishable from typical GCs in their mass and stellar populations, while others are complex objects with extended SFHs and metallicity distributions. This diversity in properties is most likely connected to how their mass assembled. A single formation scenario that explains the formation of all NSCs is incompatible with our results. In future work, we aim to go beyond the identification of the dominant NSC formation channel and estimate the relative contribution of GC-inspiral versus \textit{in-situ} formation for individual NSCs.

\begin{acknowledgements}
We thank the anonymous referee for a constructive report that helped to improve and polish this work.
GvdV acknowledges funding from the European Research Council (ERC) under the European Union's Horizon 2020 research and innovation programme under grant agreement No 724857 (Consolidator Grant ArcheoDyn). EMC is supported by MIUR grant PRIN 2017 20173ML3WW\_001 and by Padua University grants DORI1715817/17, DOR1885254/18, and DOR1935272/19. J.~F-B, FP, and IMN acknowledge support through the RAVET project by the grant PID2019-107427GB-C32 from the Spanish Ministry of Science, Innovation and Universities (MCIU), and through the IAC project TRACES which is partially supported through the state budget and the regional budget of the Consejer\'ia de Econom\'ia, Industria, Comercio y Conocimiento of the Canary Islands Autonomous Community. RMcD acknowledges financial support as a recipient of an Australian Research Council Future Fellowship (project number FT150100333). LZ acknowledges the support from National Natural Science Foundation of China under grant No. Y945271001.
\end{acknowledgements}

\bibliographystyle{aa} 
\bibliography{References}

\appendix
\section{imfit modelling of more complex galaxies}
\label{app:imfit}
\begin{figure*}
    \centering
    \includegraphics[width=0.57\textwidth]{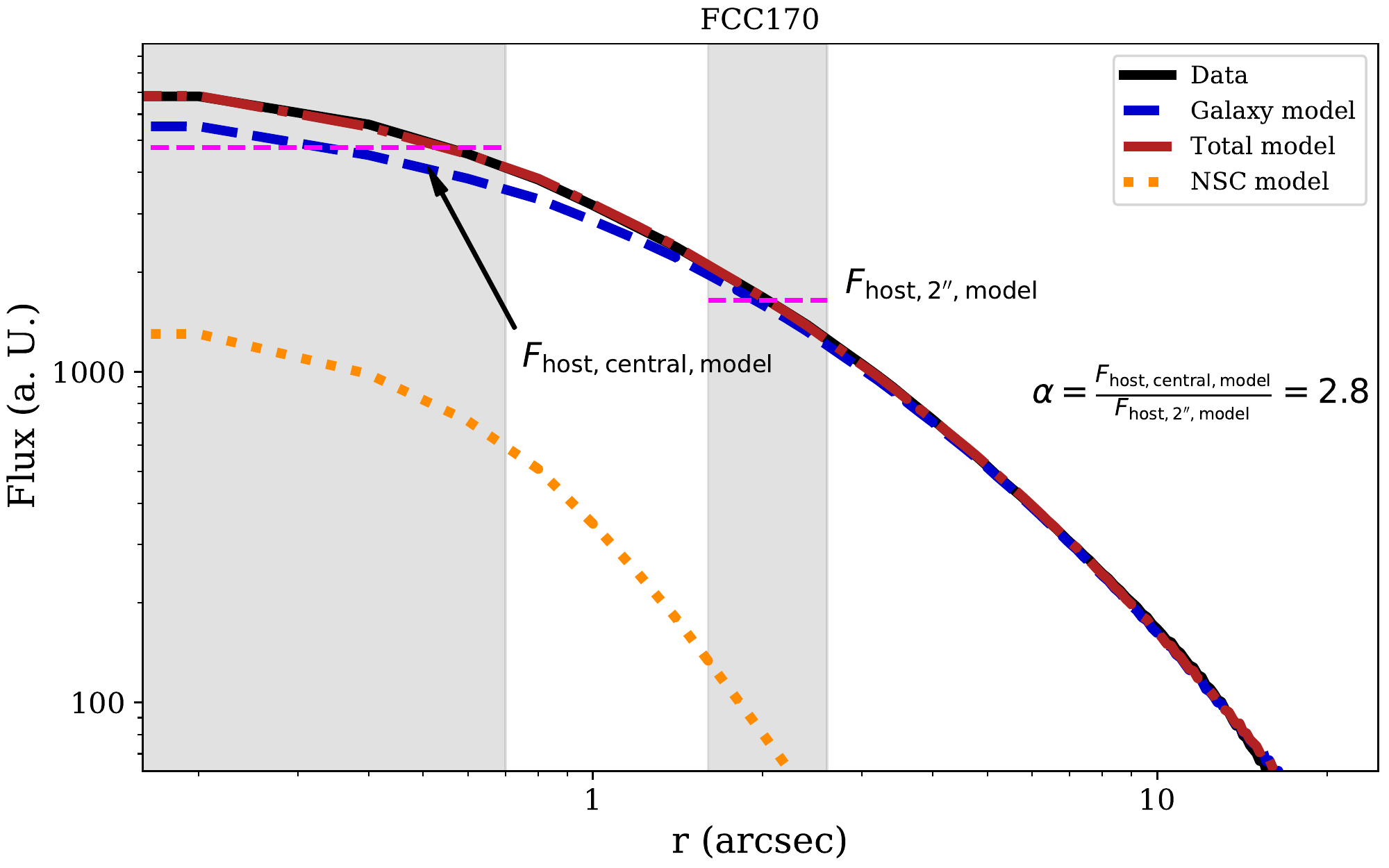}
    \includegraphics[width=0.37\textwidth]{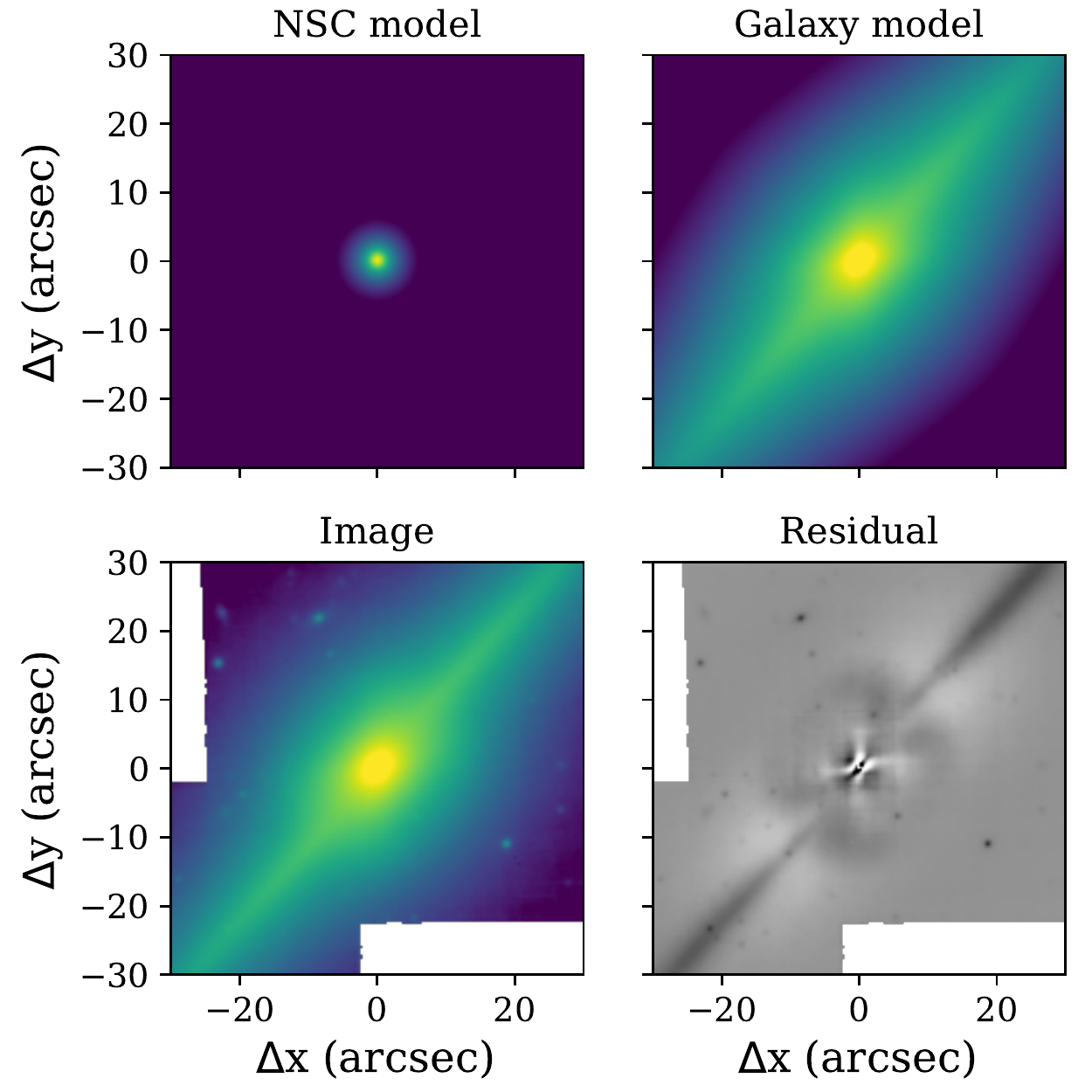}
    \caption{\textsc{imfit} modelling of FCC\,170 to decompose the white-light image into a galaxy and a NSC component, similar to Fig. \ref{fig:imfit_modelling}. \textit{Left}: radial profile of the data (solid black line), galaxy model (dashed blue line), NSC model (dotted orange line), and combined model (dashed-dotted red line). The grey areas show the extraction regions. The mean fluxes in these regions are indicated by the horizontal dashed magenta lines to illustrate the scaling factor $\alpha$. \textit{Right}: 2D cutout images of the original image (bottom left), NSC and galaxy model (top row) and resulting residual image (bottom right). In this case, the galaxy model consists of a S\'{e}rsic, a edge-on disk model and a generalised exponential function to describe the numerous components in this galaxy. The NSC model is a Moffat function. Even with this multi-component model, there are still residuals left.}
    \label{fig:imfit_modelling_FCC170}
\end{figure*}

In Sect. \ref{sect:NSC_analysis}, we describe our approach to disentangle the NSC and galaxy contribution to the central PSF spectrum based on \textsc{imfit} modelling. Fig. \ref{fig:imfit_modelling} shows this approach for one of the dwarf galaxies to illustrate the method, but especially the more massive galaxies are structurally more complex and their models still show some residuals in the 2D maps. To illustrate this, we show the decomposition of FCC\,170 in Fig. \ref{fig:imfit_modelling_FCC170}. 

FCC\,170 is an edge-on S0 galaxy with multiple structural components including a thin disk, thick disk, and a X-shaped bulge component (see e.g. \citealt{Pinna2019a, Poci2021}). Modelling this galaxy with \textsc{imfit} is challenging. Our best-fit model includes a S\'{e}rsic component, a 2D edge-on disk model, and generalised elliptical function to describe the X-shaped bulge in addition to the point-source Moffat function of the NSC\footnote{See \url{https://www.mpe.mpg.de/~erwin/code/imfit/} for a detailed description of these components}. While the radial profile exhibits a close fit of this model to the data, the 2D residual still shows structure in the centre. In this best-fit model, we find $\alpha = 2.8$, but this value is subject to uncertainties of up to $\sim 20\%$ arising from the modelling choice. For example, including further galaxy components in the centre can decrease the flux from the model NSC and thus enhance $\alpha$. For this reason, we attempted to base our set of \textsc{imfit} models on physical components in the galaxy (e.g. disks, bulge, NSC), instead of choosing an arbitrary number of components. Nonetheless, as discussed in Sect. \ref{sect:gal_specs}, our stellar population results are robust against changes in $\alpha$.

\section{Exemplary binned metallicity maps}
\label{app:maps}
\begin{figure}
    \centering
    \includegraphics[width=0.49\textwidth]{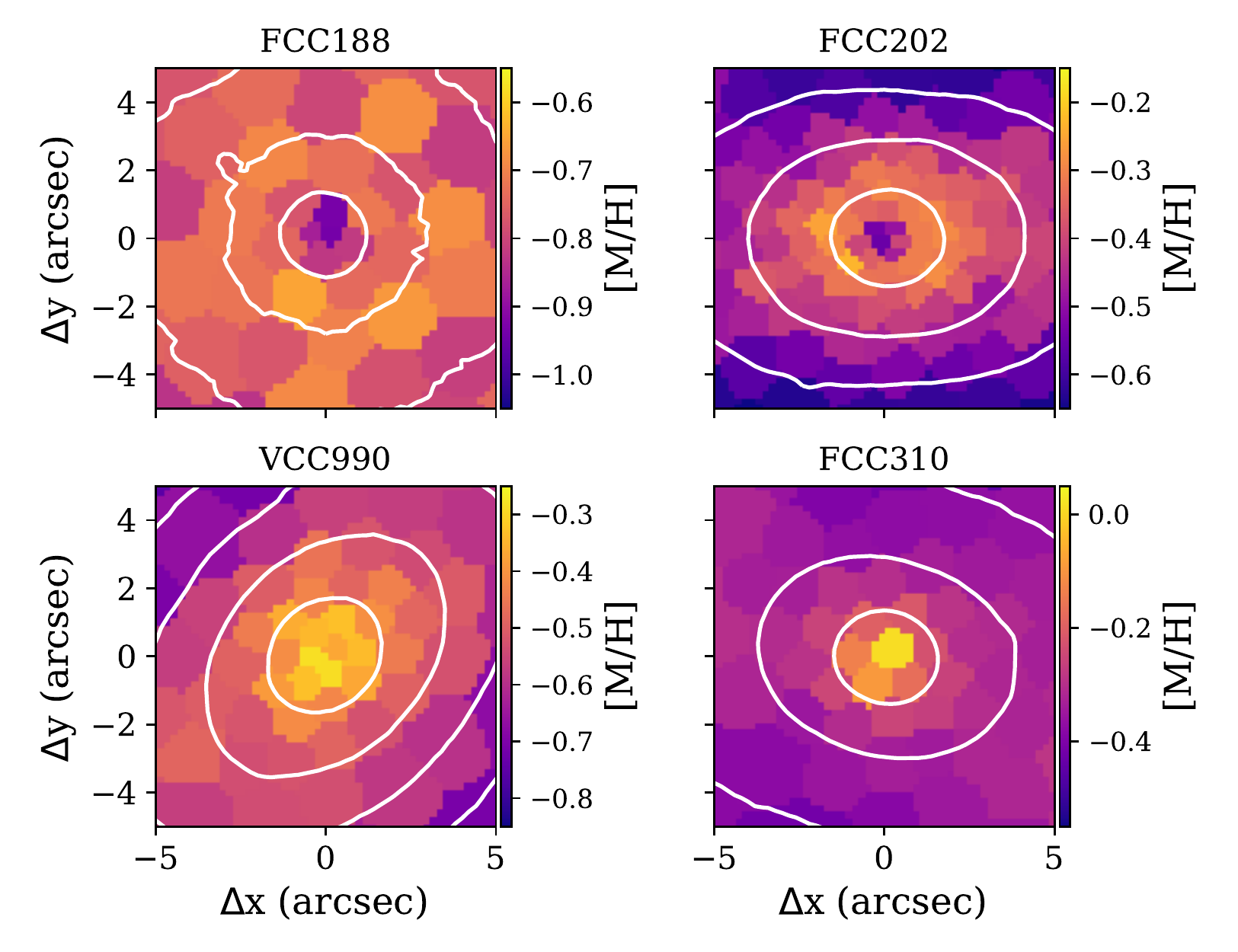}
    \caption{Voronoi-binned metallicity map of the central 5 $\times$ 5\arcsec\, of FCC\,188, FCC\,202, VCC\,990, and FCC\,310 to illustrate the central metallicity behaviour in two dimensions rather than the radial profiles presented in Fig. \ref{fig:radial_profiles}. FCC\,188 and FCC\,202 are two galaxies that show drops in metallicity at the NSC position, while VCC\,990 and FCC\,310 show increasing metallicity towards the centre. Arbitrary surface brightness levels shown in white to guide the eye. The different panels have different metallicity scaling.}
    \label{fig:maps_zoom}
\end{figure}

We present the radial metallicity profiles from Voronoi binned maps in Fig. \ref{fig:radial_profiles} and compare them to the different extraction radii. To illustrate how these one-dimensional profiles relate to the binned maps, we show cut-outs of the metallicity maps of four galaxies as examples in Fig. \ref{fig:maps_zoom}. Two of those (FCC\,188 and FCC\,202) show drops in the radial metallicity profile corresponding to central bins with lower metallicity, while the other two (VCC\,990 and FCC\,310) show higher metallicity in their centres than in the surroundings.

\section{Ages and metallicities from different approaches}
\label{app:approaches}
\begin{figure}
    \centering
    \includegraphics[width=0.49\textwidth]{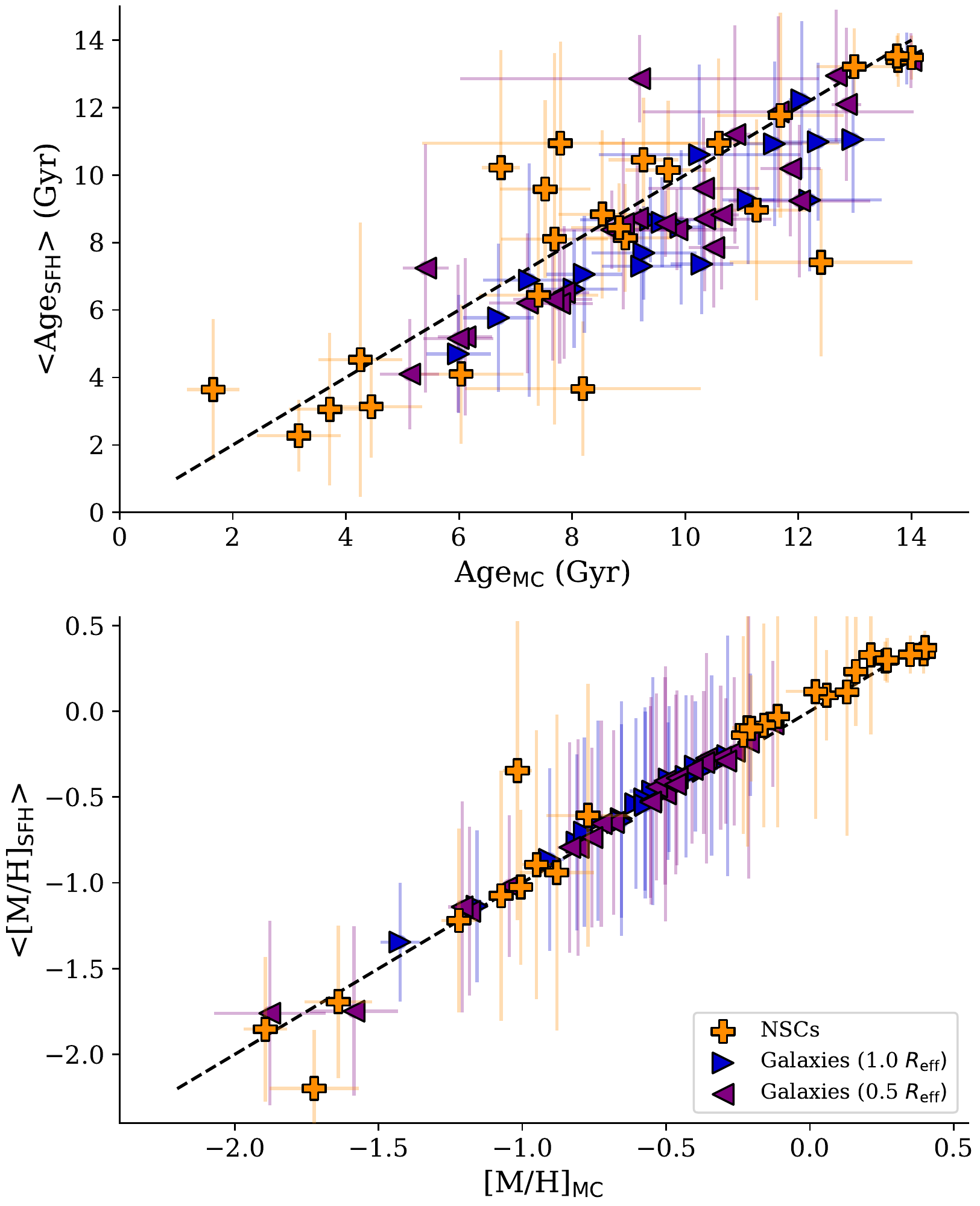}
    \caption{Comparison of mean ages (\textit{top}) and metallicities (\textit{bottom}) as inferred from the MC approach (x-axis) and regularisation approach (y-axis). For the regularisation approach, errorbars refer to the weighted uncertainties that also reflect the width of the age and metallicity distributions.}
    \label{fig:compare_approach}
\end{figure}

Figure \ref{fig:compare_approach} shows a comparison of the extracted ages and metallicities of the different components using either the weighted mean from the regularised fit or the MC approach as described in Sect. \ref{sect:ppxf}. Both methods agree within the uncertainties. The metallicities obtained from both methods agree very well, except for the NSC of FCC\,119, for which the MC approach gives a lower metallicity than the regularisation approach, although both methods agree within uncertainties. This difference is most likely caused by the presence of strong emission lines in the NSC spectrum that affect the line profiles.

\end{document}